\documentclass{aa}
\usepackage{graphicx}

\begin{document}

\title{A model of spectral galaxy evolution including the effects of nebular emission} 
\markboth{E. Zackrisson et al.: }{A model of spectral galaxy evolution including the effects of nebular emission}
\authorrunning{Zackrisson et al.}
\titlerunning{Spectral galaxy evolution including the effects of nebular emission}
\author{Erik Zackrisson\inst{1} \and Nils Bergvall\inst{1} \and Kjell Olofsson\inst{1} \and Arnaud Siebert\inst{2}}

\offprints{Erik Zackrisson, 
\email{Erik.Zackrisson@astro.uu.se}}

\institute{Astronomiska observatoriet, Box 515, S-75120 Uppsala, Sweden \and Observatoire de Strasbourg, 11 rue de l'universite, 67000 Strasbourg, France}

\date{Received 0000 / accepted 0000}

\abstract{This paper presents a new spectral evolutionary model of galaxies, properly taking the effects of nebular emission and pre-main sequence evolution into account. The impact of these features in different photometric filters is evaluated, along with the influence that variations in the physical conditions of the gas may have on broadband colours, line ratios and equivalent widths. Inclusion of nebular emission is demonstrated to radically change the predicted ultraviolet, optical and near-infrared colours during active star formation. Pre-main sequence evolution is also seen to give a non-negligible contribution to the luminosity in the near-infrared during the first few millions years of evolution and should not be omitted when very young systems are being modelled. Finally, we present a comparison of our predictions to observations and two other recent codes of evolutionary synthesis.  }

\maketitle

\section{Introduction}
Ever since the pioneering work of Tinsley (\cite{Tinsley}), spectral evolutionary models (SEMs) based on stellar population synthesis have 
proven to be important tools in the analysis of the star formation history of galaxies (e.g. Leitherer et al. \cite{Leitherer et al.1}). The Sloan survey (e.g. Gunn et al. \cite{Gunn et al.}) and the Anglo-Australian 2dF survey (e.g. Colless \cite{Colless}) will provide a vast number of galaxy spectra that, with the aid of SEMs, may be used for important statistical studies of the chemical evolution, initial mass functions and star formation histories of galaxies. 

Taking the effects of nebular emission into account becomes essential when deriving the broadband colours of galaxies subject to substantial star formation rates. This does not only apply to local starbursts, but could also have implications for various Hubble types during the first few hundred million years of evolution,  i.e. the maximum age of the most distant galaxies now observed. Despite this fact, results are continuously being published where this problem is completely overlooked, mainly due to the lack of suitable models. It is therefore a necessary and important step to include a proper treatment of nebular emission at a wide range of metallicities.

In connection with our work with metal poor galaxies we have previously developed SEMs from different approaches, all of them relevant at low metallicities and including a nebular component (Bergvall \cite{Bergvall}, Olofsson \cite{Olofsson1}). Here we present a new, more flexible and up-to-date model intended for the interpretation of low to intermediate mass galaxies within the full range of ages and metallicities observed. Improvements compared to similar codes include pre-main sequence evolution and a sophisticated treatment of nebular emission.

Since our main interest lies in the evolution of low to intermediate mass galaxies, chemical evolution is not treated at this stage. Even though a continuously increasing metallicity is definitely expected for galaxies under a closed-box scenario, most galaxies do, most certainly, undergo infall of gas. Besides hydrogen and helium, various amounts of heavy elements
should also participate in the infall process. Moreover, selective loss of chemical elements from stars within a galaxy is also expected, in particular in low-mass system (de Young et al. \cite{de Young et al.}, D'Ercole et al. \cite{D'Ercole et al.}). This effect could even be important in massive galaxies if the
star forming region resides far from the centre of the gravitational potential. Under specific conditions, the amount of heavy elements may even decrease during certain episodes in the lifetime of a galaxy (Olofsson \cite{Olofsson2}). A proper treatment of chemical evolution for low- to intermediate mass objects would require complete knowledge of the dynamics of the system, the time-scales of infall, the gas consumption rate and the ejection of chemical elements. In lack of a better prescription, we therefore assume constant metallicity evolution of each separate stellar population within a galaxy. The spectral energy distributions (SEDs) produced this way may then be mixed to simulate the effect of a metallicity distribution. We claim that this approach should be no less appropriate than closed box models subject to ad hoc modifications of the rate of infall and mass-loss necessary to match the metallicities observed. In the future we will however investigate the possibility to derive information about the chemical evolution directly from a tree-SPH model currently being developed at our institute (Pharasyn \cite{Pharasyn}), including a realistic treatment of star formation processes. 

The treatment of dust is another important but similarly difficult issue since the size, chemical composition and distribution of dust particles may vary considerably and lead to large differences in the global spectral distribution. Simple geometries of dust-star distributions will however be included in a future version of the model.

In this paper, we present the code at its present stage of development and investigate the impact of nebular emission and pre-main sequence evolution on the modelling of the SEDs of galaxies. In section 2, we outline the machinery on which the code is based - in terms of the computational method, stellar evolutionary tracks, stellar atmospheres, photoionization model and general assumptions used. In sections 3 and 4 we investigate the magnitude and duration of nebular and pre-main sequence effects in different photometric filters and for different metallicities. Section 5 includes a comparison of the colours predicted by our code to those of two other SEMs. The colours, line ratios and equivalent widths predicted are then tested against observations in section 6. Section 7 contains a summary of our findings and a few concluding remarks.

\section{The model}
The spectral evolutionary synthesis code presented in this paper uses a set of stellar evolutionary tracks, theoretical stellar atmosphere spectra and a photoionization model to predict colours and SEDs for galaxies at ages between 0-15 Gyr. The stellar component is represented by a single stellar population evolving at constant metallicity $Z$=0.001, 0.004, 0.008, 0.020 or 0.040, whose initial mass function (IMF) follows a single-valued power law
$$\Phi(m,\alpha)=\frac{\mathrm{d}N}{\mathrm{d}m}=Cm^{-\alpha}$$
within the mass interval [$m$\raisebox{-0,5ex}{l}, $m$\raisebox{-0.5ex}{u}]. The normalization constant C is here determined by the total mass converted into stars. The model allows the use of any upper and lower mass limit between 0.08-120 $M_\odot$ and any value of $\alpha$. 

The star formation history can be modelled using a constant star formation rate (SFR) of arbitrary duration ($\tau$):
$$\mathrm{SFR}(t,\tau) = \frac{M_0}{\tau} \;\;\;\;\;\;\;\;\;\; 0\le t \le \tau  $$
$$\mathrm{SFR}(t,\tau) = 0   \;\;\;\;\;\;\;\;\;\;\;\;\;\ t>\tau$$
or, alternatively, a SFR subject to exponential decrease over time: 
$$\mathrm{SFR}(t,\tau)=\frac{M_0}{\tau}\exp(-\frac{t}{\tau})$$
where $\tau$ is the e-folding decay rate of the SFR. In both cases $M_0$ represents the total mass available for star formation.

\subsection{Stellar evolutionary tracks}
In order to follow the evolution of stars between 0.08-120 $M_\odot$ at metallicities $Z$=0.001, 0.004, 0.008, 0.020 and 0.040 through all important stages of evolution, stellar tracks from several different authors have been adopted. 

The tracks developed by the Geneva group (Charbonnel et al. \cite{Charbonnel et al.1}; Schaller et al. \cite{Schaller et al.}; Schaerer et al. \cite{Schaerer et al.1}, \cite{Schaerer et al.2}; Charbonnel et al. \cite{Charbonnel et al.2}) are used to follow the evolution of stars from the zero age main sequence (ZAMS) to the end of the C-burning phase for $7\le M/M_\odot \le 120$ at all $Z$, to the end of the early asymptotic giant branch (E-AGB) for $2\le M/M_\odot \le 5$ and up to the He-flash for $0.8\le M/M_\odot < 2$ at $Z$=0.001, 0.004 and 0.008. At $Z$=0.020 and $Z$=0.040, these tracks follow the evolution to the E-AGB throughout the entire mass interval $0.8\le M/M_\odot\le 5$. Evolution from the He-flash to the E-AGB for stars between $0.8\le M/M_\odot < 2$ at $Z$=0.001, 0.004, 0.008 is instead covered using the tracks by Castellani et al. (\cite{Castellani et al.}), since the horizontal branch (HB) morphologies generated by the Geneva tracks are known to become very red regardless of metallicity. This strategy is further discussed in section 2.3.

Evolution along the pre-main sequence (PMS) has then been added for stars between $0.8\le M/M_\odot\le 7$ using the tracks by Bernasconi \cite{Bernasconi}, and from the E-AGB to the white dwarf and planetary nebula stage for stars in the range $0.8\le M/M_\odot\le 5$ using the hydrogen- and helium burning tracks by Vassiliadis \& Wood (\cite{Vassiliadis & Wood}).  In order to make the zero age post-AGB mass and main fuel a unique function of ZAMS mass, all tracks computed using artificially enhanced or reduced mass-loss rates on the red giant branch (RGB) were excluded from the set. This strategy will not affect the final predictions since the tracks actually employed cover the same range in $\log(g)$ and $T_\mathrm{eff}$ and will be interpolated during the course of calculation. 

The gap between E-AGB and the post-AGB has been bridged using characteristic  luminosities and time-scales ($\sim$ 10$^5$ yr) of the thermally pulsing AGB phase (Willson \cite{Willson1}, \cite{Willson2}) for all stars 
less massive than 5 $M_\odot$. These numbers are based on a combination of theoretical mass loss rates and observations, in which selection effects have carefully been taken into account.

Low-mass tracks from the Lyon-group (Chabrier \& Baraffe \cite{Chabrier & Baraffe}) have been adopted to encompass the evolution of stars with masses in the range $0.08\le M/M_\odot < 0.8$. Since these tracks terminate at an age of 10 Gyr, it has been necessary to make linear extrapolations in $\log(t)$ in order to obtain a final data point at an age of 15 Gyr. These models use non-grey atmospheres, which resolves the temperature-luminosity discrepancies at the bottom of the low-mass sequence present in previous generations of low-mass models. Kroupa \& Tout (\cite{Kroupa & Tout}) compared several low-mass models and concluded that the ones created by the Lyon-group provided the best agreement with observations.

All tracks not available at the metallicities used in the Geneva set have been been linearly inter- and extrapolated in $\log(Z)$ to cover all five metallicities chosen for our model, and, in the case of the post-AGB tracks, also to cover the full mass range at $Z$=0.001. The ZAMS to E-AGB tracks by Charbonnel et al. \cite{Charbonnel et al.2} have been extrapolated up to $Z$=0.040 using a procedure that closely matches each track to the original ZAMS to He-flash tracks by Schaerer et al. (\cite{Schaerer et al.1}) during common stages of evolution.

\subsection{Important approximation connected to PMS evolution}
PMS stars are assumed not to contribute to the total luminosity of the galaxy before emerging from their dusty parental cocoons. The time-scales of these processes have been adopted from Bernasconi (\cite{Bernasconi}). Even though this correction is probably appropriate in the optical region, as illustrated by the gap between the Hayashi track and the observed birth line of T Tauri stars in the Hertzsprung-Russel diagram (e.g Stahler \cite{Stahler}), it may not be as justified in the infrared (IR). Including these embedded stars would however require substantial effort, since their SEDs are broader than a single temperature blackbody (Shu et al. \cite{Shu et al.}) and may not be well-represented by our current library of stellar atmospheres. 

The correction for the parental cocoon only applies to the pre-main sequence tracks used in the range $0.8\le M/M_\odot \le 7$, and not to the low-mass tracks $0.08\le M/M_\odot < 0.8$ (covering the pre-main sequence as well as later stages) due to the lack of detailed pre-main sequence time-scales for these objects. 

\subsection{Horizontal branch morphologies}
Since the process responsible for mass-loss on the red giant branch (RGB) is not properly understood, HB models are still highly uncertain. This becomes especially apparent in low metallicity globular clusters (GCs), where standard models are unable to reproduce the wide range of HB morphologies observed. In order to set the stage for our approach to this problem, we present a short review of the on-going debate about the second parameter (metallicity being the first) responsible for determining the morphology.   
	
By assuming that more massive stars experience more intense mass loss along the RGB, age becomes a natural candidate for the second parameter (e.g. Fusi-Pecci \& Renzini \cite{Fusi}). However, as has been argued by Buonanno
et al. (\cite{Buonanno}) several parameters may play more or less
significant role both within a specific cluster and between clusters.
As was discussed by Catalan \& de Freitas Pacheco (\cite{Catelan1}), the two
clusters M3 and M13 which have a similar age (Catelan \& de Freitas Pacheco \cite{Catelan1}; Johnson \& Bolte \cite{Johnson}; VandenBerg \cite{VandenBerg1}; Grundahl \cite{Grundahl}) still show a large difference in HB morphology. Similarly, Catelan (\cite{Catelan2}) and VandenBerg 
(\cite{VandenBerg2}), found that the difference in HB morphology between other
galactic GCs proved too large to be accounted for by age dependent mass
loss alone and that age could {\it not} be the dominant secondary parameter. 

It has been known for a long time that the Sculptor
dwarf galaxy contains an unusually red HB population for its
high age and low metallicity (Zinn \cite{Zinn}; DaCosta \cite{DaCosta}; Mateo \cite{Mateo}). In
their recent study of the galaxy Hurley-Keller et al. (\cite{Hurley}) note 
that while there is a strong concentration of these red HB stars
towards the centre, no gradient in age is found.

There seems
to be a systematic tendency for GCs with blue extensions of the HB to contain
super oxygen-poor stars, as measured by the [O/Fe] relative abundances (Catelan \& de Freitas Pacheco \cite{Catelan2}; Carretta \& Gratton \cite{Carretta & Gratton}). Since stars with these chemical properties appear to be losing mass at an abnormally high rate, affecting RGB evolution as well as later stages, detailed differences in chemical composition may well contribute to the second
parameter effect. It is therefore necessary to understand the mechanism
that influences the variations of chemical composition of the stellar
atmospheres. Cottrell \& da Costa (\cite{Cottrell}) suggested a primordial
origin of the variations, possibly originating in the processed envelopes of intermediate-mass AGB
stars that polluted the nearby cluster environment by different amounts. In
this case the chemical variations of the cluster RGB stars would be of a purely
stochastical nature. Peterson (\cite{Peterson}) suggested that the second parameter might be related to stellar rotation. If so, such an effect may also explain the
abundance anomalies through the induction of deep mixing processes. Most likely
there would be a combination of primordial and induced origin.

Another  possibility is that the RGB stars are spun up by stellar
companions resulting in an increased mass loss and a subsequent tendency to
end up in the blue part of the HB. Rich (\cite{Rich}), however, regard the influence
from binary companions to be insufficient. More importantly, and possibly
correlated with the relative frequency of binaries and thus complementary, may
be the influence of massive planets as argued by Soker \& Harpaz (\cite{Soker}).
 
The conclusion must be that there may be a rich spectrum of factors behind the
second parameter effect and that age probably plays a minor role. Even so, most SEMs generally employ evolutionary tracks from either the Geneva or Padova group without further modifications, and therefore implicitly assume the second parameter to be age. To study how this hypothesis may affect the predicted spectral evolution of galaxies we have chosen to adopt an opposing view. 

In order to properly cover the whole range of observed morphologies at low metallicities, the theoretical HB-AGB tracks by Castellani et al. (\cite{Castellani et al.}) have been compared to the morphological classification of HBs for a large number of Galactic GCs (Harris \cite{Harris}). For $Z$=0.001, 0.004 and 0.008, all tracks crossing the instability strip in a way that does not seem to be supported by these observations have been excluded from the set.  Every star with initial mass below 1.7 $M_\odot$ is then assumed to be equally likely to follow any of the remaining HB-AGB tracks. This procedure will simulate a second parameter that is stochastic in nature and result in an age-independent distribution of stars along the HB that we claim will give a better coverage of the many HB morphologies (especially at $Z$=0.001) confirmed observationally.

Due to the scarce observational material on higher metallicity GCs, this technique breaks down at $Z$=0.020 and $Z$=0.040, for which we have chosen to adopt the more mainstream approach provided by the use of Geneva tracks (Charbonnel et al. \cite{Charbonnel et al.2}). This should not result in any great inconsistency in terms of morphology, unless the HB becomes progressively bluer at metallicities above $Z$=0.008.

Since our HB strategy assumes the existence of a large variety of horizontal branch morphologies within each system at low metallicities, caution must be taken when applying the model to very small stellar systems (e.g. individual globular clusters). It should of course be noted that if the second parameter truly is stochastic in nature, the uncertainties introduced by doing so may perhaps be no larger than those introduced when using conventional models.

\subsection{Theoretical stellar atmospheres}
The compilation of theoretical stellar atmospheres by Lejeune et al. (\cite{Lejeune et al.}) has been used for stars with effective temperature $\le$ 40000 K. These cover effective temperatures as low as 2000 K, surface gravities in the range $-1.02 \le \log(g) \le 5.5$ and metallicities $-3.5 \le [M/\mathrm{H}] \le 1.0$. 

In order to match the five metallicities covered by our stellar evolutionary tracks, the original grid has been linearly interpolated in $\log(Z)$. 
For effective temperatures over 40000 K, the non-LTE atmospheres by Clegg \& Middlemass (\cite{Clegg & Middlemass}) have been employed after linear interpolation between data points to yield the same resolution as for the Lejeune compilation (1221 data points between wavelengths 9.1-160000 nm). 

The two sets of theoretical atmospheres are not completely compatible, and making the switch at 40000 K will cause an unavoidable discontinuity in the ultraviolet (UV) spectral evolution. This will manifest itself in a sudden drop of several orders of magnitude in flux at wavelengths below 23 nm as the stars of the stellar population cool off sufficiently to fall below this temperature. 

\subsection{The nebular component}
The inclusion of a nebular component is of vital importance for modelling regions of active star formation. Even so, many recent models choose to neglect the effect of ionized gas or to treat it only superficially. The Starburst99 model of Leitherer et al. (\cite{Leitherer et al.2}) includes the nebular continuum, but no emission lines. The PEGASE2 model of Fioc and Rocca-Volmerange (\cite{Fioc & Rocca-Volmerange2}) predicts a gas continuum as well as emission lines, but only by resorting to a nebular component pre-calculated for a set of fixed gas parameters\footnote{However, a coupling of the two codes PEGASE2 and Cloudy was recently published by Moy et al. (\cite{Moy et al.}) and applied to starbursts and HII galaxies.}. 

Our nebular component (lines and continuum) is modelled using the predicted SED of the stellar population as a heat source for the photoionization code Cloudy version 90.05 (Ferland \cite{Ferland}). This makes our gas prescription far more flexible than existing models, and allows the possible impact of variations in the physical conditions of the gas on spectral evolution to be evaluated. Variable input parameters include the covering factor, filling factor, gas metallicity and hydrogen number density. The covering factor simulates the existence of possible holes in the nebula, whereas the filling factor quantifies the degree of graininess of the gas.

When applying Cloudy, some important approximations have been used: The stars are assumed not to be spatially mixed with the gas, the hydrogen density is assumed to be constant in time, the nebula to be ionization bounded and the geometry to be spherical.

\subsection{Numerical method}
Following Fioc \& Rocca-Volmerange (\cite{Fioc & Rocca-Volmerange}), the monochromatic flux of a galaxy at time $t_{\mathrm{gal}}$ may be written:
{\setlength\arraycolsep{-30pt}\begin{eqnarray}
F_{\lambda}(t_{\mathrm{gal}},\tau,\alpha)=\int_{0}^{t_{\mathrm{gal}}}\int_{m_{\mathrm{l}}}^{m_{\mathrm{u}}} \mathrm{SFR}(t_{\mathrm{gal}}-t_{\mathrm{star}},\tau) \Phi(m,\alpha)\cdot \nonumber\\ & & \cdot f_{\lambda}(m,t_{\mathrm{star}}) \ \mathrm{d}m \ \mathrm{d}t_{\mathrm{star}} \nonumber
\end{eqnarray}} 
provided that $\Phi(m,\alpha)$ is normalized to 1 $M_\odot$. Here, $f_{\lambda}(m,t_{\mathrm{star}})$ represents the monochromatic flux of a star with initial mass $m$ at wavelength $\lambda$ and age $t_{\mathrm{star}}$. 

In order to perform spectral evolutionary synthesis, at least some part of this expression is commonly discretized. Discretizing both integrals leads to a somewhat outdated method of evolutionary synthesis, whose drawbacks were thoroughly investigated by Charlot \& Bruzual (\cite{Charlot & Bruzual}), whereas discretization of only one integral leads to the more modern  isochrone (time integral discretized) and isomass (mass integral discretized) methods. This model is based on the latter technique, where the expression for $F_{\lambda}(t_{\mathrm{gal}},\tau,\alpha)$ is rewritten as:
{\setlength\arraycolsep{-40pt}\begin{eqnarray}
F_{\lambda}(t_{\mathrm{gal}},\tau,\alpha)=\sum_{i=1}^{p-1} \mathrm{SFR}(t_{\mathrm{gal}}-t_{\mathrm{star},i},\tau) \cdot \nonumber\\ & & \cdot \sum_{j=1}^{q-1} \Phi(m_{\mathrm{j}},\alpha) (m_{\mathrm{j+1}}-m_{\mathrm{j}})\int_{t_{\mathrm{star},i}}^{t_{\mathrm{star},i+1}} f_{\lambda}(m_{\mathrm{j}},t_{\mathrm{star}}) \ \mathrm{d}t_{\mathrm{star}}\nonumber
\end{eqnarray}}
Here $t_{\mathrm{star},1}=0$, $t_{\mathrm{star},p}=t_{\mathrm{gal}}$, $m_{\mathrm{1}}=m_{\mathrm{l}}$ and $m_{\mathrm{q}}=m_{\mathrm{u}}$. The number of time and mass intervals used in the evaluation is represented by $p-1$ and $q-1$, respectively. 

The problem of performing a simultaneous discretization of both time and mass integrals is that it may cause oscillating colours and luminosities when modelling short bursts of star formation. These phenomena arise when stars on one track abruptly move into a different evolutionary phase that stars on the consecutive track will not reach until later. Oscillations of this kind were a major reason for Charlot \& Bruzual (\cite{Charlot & Bruzual}) to propose the technique of isochrone synthesis as a better alternative, although the isomass method should theoretically be just as valid. This was pointed out by Fioc \& Rocca-Volmerange (\cite{Fioc & Rocca-Volmerange}) who also claimed to have tested both methods with identical results. There is, however, always the risk that hidden approximations present in the implementation of these algorithms may have an impact under certain circumstances. Since our ambition has been to develop an independent code that allows critical evaluation of the uncertainties involved in the predictions, we have therefore preferred the use of the isomass method to the more mainstream isochrone approach.

In this technique, it is necessary to ensure that $m_{\mathrm{j+1}}$-$m_{\mathrm{j}}$ is sufficiently small so that equivalent evolutionary phases of consecutive masses overlap. For this reason, the original mass grid is interpolated to yield a set of stellar evolutionary tracks appropriate for each metallicity and set of time steps. 

\section{The impact of nebular emission}
\begin{table}[t]
\caption[]{Duration of measurable contributions in Johnson filters from nebular emission. The contributions have been evaluated at metallicities $Z$=0.001 and $Z$=0.020 for three different star formation scenarios: A short burst of constant star formation ($\tau$=100 Myr) (S) and exponentially decreasing SFRs with e-folding decay rates $\tau$=1 Gyr (E1) and $\tau$=14 Gyr (E14).
All scenarios were computed using a Salpeter IMF over the mass range 0.08-120 $M_\odot$, $10^{10} \ M_\odot$ available for star formation, $Z_\mathrm{stars}$=$Z_\mathrm{gas}$, covering factor 1.0, filling factor 1.0 and $n$(H)=100 $\mathrm{cm^{-3}}$. The entries $\tau$(filter) measure the time in Myr after which the discrepancy between models with and without the nebular component become negligible, i.e. smaller than the typical observational uncertainty of 0.05 magnitudes.}

\begin{flushleft}
\begin{tabular}{lllllll} 
\hline
	 & $Z$=0.001 & & & $Z$=0.020 & & \cr
	 & S  & E1 & E14 & S & E1 & E14 \cr
\hline 
	  $\tau$(U) & 100 & 3000 & 15000 & 100 & 3000 & 15000\cr
	  $\tau$(B) & 100 & 3000 & 15000 & 100 & 2000 & 15000 \cr
	  $\tau$(V) & 100 & 3000 & 15000 & 100 & 2000 & 12000 \cr
	  $\tau$(R) & 100 & 2000 & 14000 & 100 & 1000 & 9000\cr
	  $\tau$(I) & 100 & 1000 & 2000 & 100 & 1000 & 900\cr
	  $\tau$(J) & 100 & 1000 & 4000 & 100 & 500 & 900\cr
	  $\tau$(K) & 100 & 1000 & 5000 & 100 & 800 & 1000\cr
\hline
\end{tabular}
\end{flushleft}
\end{table}

\begin{figure}[t]
\resizebox{\hsize}{!}{\includegraphics{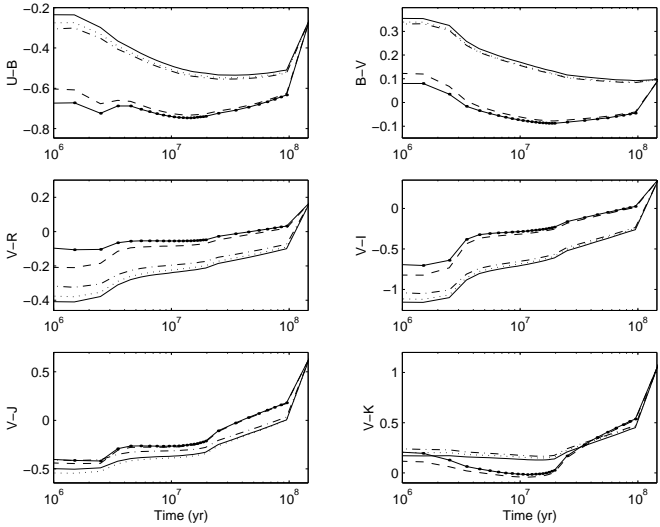}}
\caption[]{The evolution of the $U-B$, $B-V$, $V-R$, $V-I$, $V-J$ and $V-K$ colours predicted by our model at $Z$=0.001 when the covering factor, filling factor and hydrogen number density ($\mathrm{cm^{-3}}$) are allowed to take on the following values: 1.0, 1.0, 100 (solid); 0.5, 1.0, 100 (dashed); 1.0, 0.1, 100 (dash-dotted); 1.0, 1.0, 10 (dotted); 0.5, 0.1, 10 (solid with dots). All evolutionary sequences assume a Salpeter IMF (0.08-120 $M_\odot$), a short burst of constant star formation ($\tau$=100 Myr), $10^{10} \ M_\odot$ available for star formation and $Z_{\mathrm{stars}}$=$Z_{\mathrm{gas}}$. }
\label{nebcol_paramtest_z001}
\end{figure}
We here investigate the magnitude and duration of nebular effects on broadband colours, and test the sensitivity of colours and emission line equivalent widths to the metallicity and physical conditions of the gas. 

In Table 1, the duration of observable nebular effects in different photometric filters is calculated for two different metallicities and three star formation scenarios. As seen, inclusion of the nebular component may alter the photometric properties of galaxies from the UV to the near-IR during substantial periods of time. These time-scales may serve as estimates to the age of a galaxy after which nebular effects may be neglected, a point often overlooked in the analysis of high-redshift objects. For each filter, the time-scale of significant nebular emissions is determined by the temporal evolution of the bright emission lines and the relative importance of continuum emission. For instance, the sharp decline in the time-scales between filters R and I in the most prolonged star formation scenario is caused by the presence of H$\alpha$ in the R-band and the lack of durable lines of sufficient luminosity in I. The subsequent increase in time-scales towards longer wavelengths is due to increasing continuum emission throughout filters I, J and K.

\begin{figure}[t]
\resizebox{\hsize}{!}{\includegraphics{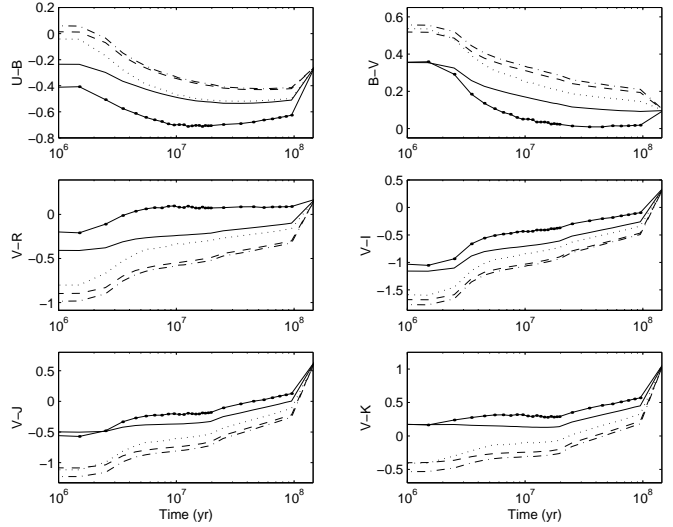}}
\caption[]{The evolution of $U-B$, $B-V$, $V-R$, $V-I$, $V-J$ and $V-K$ colours predicted by our model for a stellar metallicity $Z_{\mathrm{stars}}$=0.001 and gas metallicities $Z_{\mathrm{gas}}$=0.001 (solid), $Z_{\mathrm{gas}}$=0.004 (dashed), $Z_{\mathrm{gas}}$=0.008 (dash-dotted), $Z_{\mathrm{gas}}$=0.020 (dotted), $Z_{\mathrm{gas}}$=0.040 (solid with dots), assuming covering factor 1.0, filling factor 1.0, hydrogen number density 100 $\mathrm{cm^{-3}}$ and the same IMF and star formation history as in Fig.~\ref{nebcol_paramtest_z001}.}
\label{nebcol_ztest}
\end{figure}
\begin{figure}[t]
\resizebox{\hsize}{!}{\includegraphics{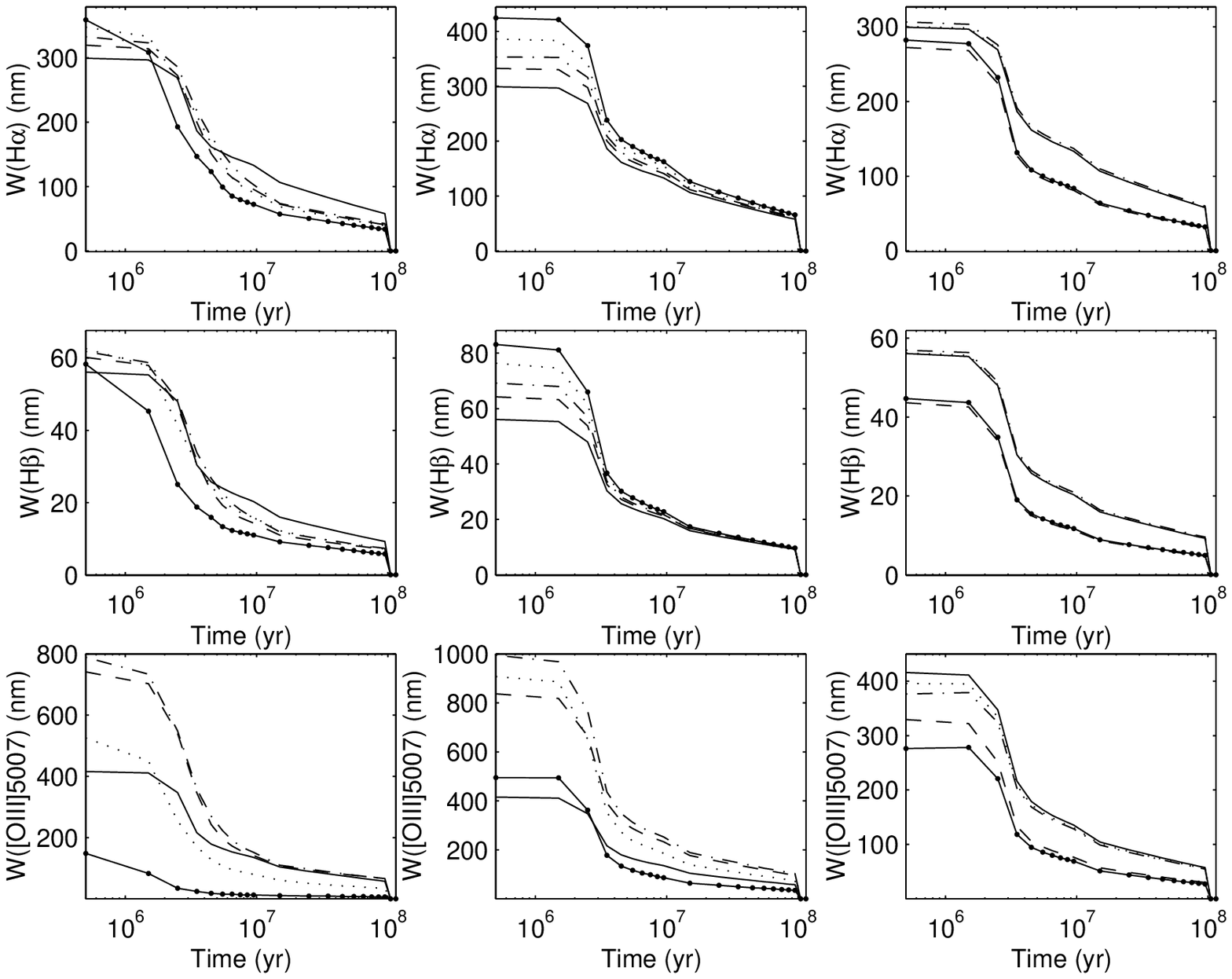}}
\caption[]{The evolution of the H$\alpha$,  H$\beta$ and 
[OIII]$\lambda$5007 equivalent widths for the same IMF and star formation history as in Fig.~\ref{nebcol_paramtest_z001}. Left column: Metallicities ($Z_{\mathrm{stars}}$=$Z_{\mathrm{gas}}$) $Z$=0.001 (solid), $Z$=0.004 (dashed), $Z$=0.008 (dash-dotted), $Z$=0.020 (dotted), $Z$=0.040 (solid with dots). Middle column: Stellar metallicity $Z_{\mathrm{stars}}$=0.001 and gas metallicities $Z_{\mathrm{gas}}$=0.001 (solid), $Z_{\mathrm{gas}}$=0.004 (dashed), $Z_{\mathrm{gas}}$=0.008 (dash-dotted), $Z_{\mathrm{gas}}$=0.020 (dotted), $Z_{\mathrm{gas}}$=0.040 (solid with dots), Right column: Metallicity ($Z_{\mathrm{stars}}$=$Z_{\mathrm{gas}}$) $Z$=0.001, covering factor, filling factor and hydrogen number density ($\mathrm{cm^{-3}}$) allowed to take on the following values: 1.0, 1.0, 100 (solid); 0.5, 1.0, 100 (dashed); 1.0, 0.1, 100 (dash-dotted); 1.0, 1.0, 10 (dotted); 0.5, 0.1, 10 (solid with dots).}
\label{EW}
\end{figure}
By varying the parameters regulating the physical properties of the gas we have also investigated the sensitivity of the predicted colours to such factors. The sensitivity to reasonable variations of the covering factor, filling factor and hydrogen number density for a short burst scenario is illustrated in Fig.~\ref{nebcol_paramtest_z001}. As expected, the most dramatic effects come  from variations in the covering factor, shifting the colours towards those of the a naked stellar component for low parameter values. Although the effective value of this parameter remains unknown, it is fair to say that no studies to date have unraveled large Lyman continuum escape fractions (which would be one possible signature of significant departures from covering factor unity) from galaxies in the local universe\footnote{But see the recent study by Steidel et al. (\cite{Steidel et al.}) for a different picture at high redshift.}. More troublesome is perhaps the fact that variations in filling factor and hydrogen density may alter the colours by as much as 0.1 magnitudes.
The dependence of predicted colours on gas metallicity, when the stellar metallicity is held constant, is plotted in Fig.~\ref{nebcol_ztest}. Here, the dependence is shown to be even stronger than for the covering factor, peaking at 0.8 magnitudes in $V-R$. Since even modest variations of the physical properties of the gas may have such a significant impact on the predictions, the common procedure of assuming parameter values typical of star-forming regions without a more careful analysis of the possible parameter space should be considered quite a hazardous exercise.

In Fig.~\ref{EW}, we present the evolution of the H$\alpha$, H$\beta$ and 
[OIII]$\lambda$5007 line equivalent widths for different combinations of gas parameters and metallicities. When the stellar and nebular metallicities are considered to have the same value, the H$\alpha$ and H$\beta$ equivalent widths only show a modest metallicity dependence, whereas the metallicity dependence of [OIII]$\lambda$5007 is rather pronounced. In the case when the stellar metallicity is held constant at $Z_{\mathrm{stars}}$=0.001 and the gas metallicity $Z_{\mathrm{gas}}$ is allowed to vary, both H$\alpha$ and H$\beta$ equivalent widths increase with increasing  $Z_{\mathrm{gas}}$. For the [OIII]$\lambda$5007 line, the highest equivalent widths are instead produced by the intermediate $Z_{\mathrm{gas}}$. The H$\alpha$ and H$\beta$ equivalent widths are apparently only sensitive to the covering factor, whereas the [OIII]$\lambda$5007 line also shows a slight dependence on filling factor and density. 

Stasi\'nska \& Leitherer (\cite{Stasinska & Leitherer}) investigated the evolution of the [OIII]$\lambda$5007 line equivalent width in a similar way, predicting monotonically decreasing equivalent widths for lower metallicities, but oscillating and increasing widths at solar metallicity, thereby hampering the use of this line as an age indicator for high-metallicity objects. This behavior is not reproduced by our model, where the width shows a monotonic decrease with age for all metallicities, although not a very strong one at $Z$=0.040. The origin of this discrepancy is not known, but could possibly be connected to the lack of realistic atmosphere spectra for Wolf-Rayet stars in the library of Clegg \& Middlemass (\cite{Clegg & Middlemass}).

\section{The impact of pre-main sequence evolution}
\begin{table}[h]
\caption[]{Duration and amplitude of measurable contributions in Johnson filters from the PMS tracks by Bernasconi \cite{Bernasconi}, evaluated by comparing evolutionary sequences with and without the inclusion of the light emitted during this stage. The entries max($\Delta$(filter)) measure the maximum difference in magnitudes between models with and without PMS light (negative max($\Delta$(filter)) indicate an increase in luminosity due to the inclusion of light emitted during this stage). The entries $\tau$(filter) measure the time in Myr after which the discrepancy between models with and without pre-main sequence light become negligible, e.g. smaller than the typical observational uncertainty of 0.05 magnitudes.}
\begin{flushleft}
\begin{tabular}{lllll} 
\hline
	 & $Z$=0.001    &         & $Z$=0.020 & \cr
	 & Standard   &  Extreme & Standard & Extreme\cr
\hline
          max($\Delta U$) & -0.056 & -0.293 & -0.036 & -0.169\cr
	  max($\Delta B$) & -0.108 & -0.507 & -0.065 & -0.292\cr
	  max($\Delta V$) & -0.153 & -0.669 & -0.089 & -0.407\cr
	  max($\Delta R$) & -0.200 & -0.767 & -0.118 & -0.527\cr
	  max($\Delta I$) & -0.270 & -0.814 & -0.163 & -0.599\cr
	  max($\Delta J$) & -0.402 & -0.877 & -0.256 & -0.627\cr
	  max($\Delta K$) & -0.561 & -0.814 & -0.391 & -0.617\cr
	  $\tau$(U) & 1 & 13 & 0 & 7 \cr
	  $\tau$(B) & 2 & 31 & 1 & 24 \cr
	  $\tau$(V) & 3 & 43 & 2 & 40 \cr
	  $\tau$(R) & 3 & 51 & 3 & 46 \cr
	  $\tau$(I) & 3 & 56 & 3 & 35 \cr
	  $\tau$(J) & 4 & 59 & 4 & 15 \cr
	  $\tau$(K) & 17 & 53 & 5 & 10 \cr
\hline
\end{tabular}
\end{flushleft}
\end{table}
\begin{table}[h]
\caption[]{Duration and amplitude of measurable contributions in Johnson filters from the PMS tracks by Bernasconi \cite{Bernasconi}, evaluated by comparing evolutionary sequences including these tracks to those assuming all stars to start on the ZAMS $t$=0. The entries max($\Delta$(filter)) measure the maximum difference in magnitudes between models with and without PMS tracks (negative max($\Delta$(filter)) indicate an increase in luminosity due to the inclusion of this phase). Otherwise same as Table 2.}
\begin{flushleft}
\begin{tabular}{lllll} 
\hline
	 & $Z$=0.001    &         & $Z$=0.020 & \cr
	 & Standard   &  Extreme & Standard & Extreme\cr
\hline
          max($\Delta U$) & 0.030 & 0.111 & 0.047 & 0.196\cr
	  max($\Delta B$) & 0.025 & 0.070 & 0.068 & 0.256\cr
	  max($\Delta V$) & 0.012 & -0.100 & 0.066 & 0.217\cr
	  max($\Delta R$) & -0.041 & -0.212 & 0.052 & 0.130\cr
	  max($\Delta I$) & -0.089 & -0.326 & 0.027 & 0.029\cr
	  max($\Delta J$) & -0.191 & -0.477 & -0.040 & -0.160\cr
	  max($\Delta K$) & -0.340 & -0.566 & -0.157 & -0.301\cr
	  $\tau$(U) & 0 & 1 & 0 & 6 \cr
	  $\tau$(B) & 0 & 1 & 1 & 9 \cr
	  $\tau$(V) & 0 & 8 & 1 & 8 \cr
	  $\tau$(R) & 0 & 16 & 1 & 7 \cr
	  $\tau$(I) & 2 & 25 & 0 & 0 \cr
	  $\tau$(J) & 3 & 34 & 0 & 6 \cr
	  $\tau$(K) & 11 & 42 & 3 & 6 \cr
\hline
\end{tabular}
\end{flushleft}
\end{table}
Despite sporadic evidence (e.g. Charlot \cite{Charlot}) pointing in a different direction, the PMS is normally assumed to be too short-lived or faint to seriously affect the integrated SED of stellar populations. By comparing evolutionary scenarios with and without inclusion of this phase, we have tested the validity of this assumption. The results are presented in Tables 2 and 3, where the contribution to the integrated luminosity from the PMS tracks by Bernasconi \cite{Bernasconi}, subject to the approximations specified in section 2.2, has been evaluated in different photometric filters. Table 2 compares scenarios which invoke these tracks to cover the time gap in the Geneva set from the Hayashi-track to the ZAMS, with and without the inclusion of the light emitted during this phase. Table 3 instead compares scenarios which include the light emitted and time spent on the Bernasconi tracks to those where all stars are assumed to climb the ZAMS at time equal to zero. Which of these tables that best describe the PMS contribution to the SED depends on how the time zero is defined in its absence. 

Both tables evaluate the PMS contributions at metallicities $Z$=0.001 and $Z$=0.020 for two different IMFs: A standard scenario computed using a Salpeter IMF over the mass range 0.08-120 $M_\odot$, and an extreme scenario, designed to maximize the effects of the lowest mass stars using a different IMF ($\alpha$=-2.85) and the smaller mass range 0.08-25 $M_\odot$. Both scenarios were calculated using a short burst of star formation ($\tau$=100 Myr) and time steps of 1 Myr.

Regardless of evaluation method, the PMS can be seen to produce measurable signatures in the near-IR evolution during the first few Myr of a standard IMF, short burst scenario. If more extreme IMFs are to be modelled, the effects become significant in all filters and for much more prolonged periods of time. The PMS should therefore certainly not be neglected when very young stellar populations are being modelled. 
\begin{table*}[t]
\caption[]{Time-weighted averages of the colour discrepancies between the evolutionary sequences from our model (Z2001), PEGASE2 (PEG2) and BC96 seen in the comparison of Fig.~\ref{galpegbc_s}. Each column represents the comparison between two sequences.}
\begin{flushleft}    
\begin{tabular}{llllllllll}
\hline
& Z2001 & Z2001 & PEG2  & Z2001 & Z2001 & PEG2  & Z2001 & Z2001 & PEG2    \cr
& PEG2  & BC96     & BC96     & PEG2  & BC96     & BC96     & PEG2     & BC96  & BC96 \cr
& $Z$=0.004  & $Z$=0.004  & $Z$=0.004  & $Z$=0.008  & $Z$=0.008 & $Z$=0.008 & $Z$=0.020 & $Z$=0.020 & $Z$=0.020 \cr
\hline
          $<|\Delta(U-B)|>$& 0.056 & 0.116 & 0.071 & 0.030 & 0.074 & 0.050 & 0.045 & 0.040 & 0.087\cr
          $<|\Delta(B-V)|>$& 0.097 & 0.097 & 0.011 & 0.013 & 0.051 & 0.013 & 0.024 & 0.026 & 0.008\cr
	  $<|\Delta(V-K)|>$& 0.305 & 0.313 & 0.032 & 0.346 & 0.334 & 0.038 & 0.255 & 0.258 & 0.055\cr
\hline
\end{tabular} 
\end{flushleft}
\end{table*}

\section{Comparison to other models}
To illustrate the uncertainties still present in contemporary spectral evolutionary synthesis, we present a comparison between colour predictions from our model and those produced by the widely-used codes of Bruzual \& Charlot (\cite{Bruzual & Charlot}) (hereafter BC96) and Fioc \& Rocca-Volmerange (\cite{Fioc & Rocca-Volmerange2}) (PEGASE2) when similar input parameters are being used. Johnson filters have been employed throughout. 

PEGASE2 enables a comparison between stellar and nebular components at all five metallicities used by our model, whereas BC96 only allows comparison between stellar components at $Z$=0.004, 0.008 and 0.020. 

\begin{figure}[h]
\resizebox{\hsize}{!}{\includegraphics{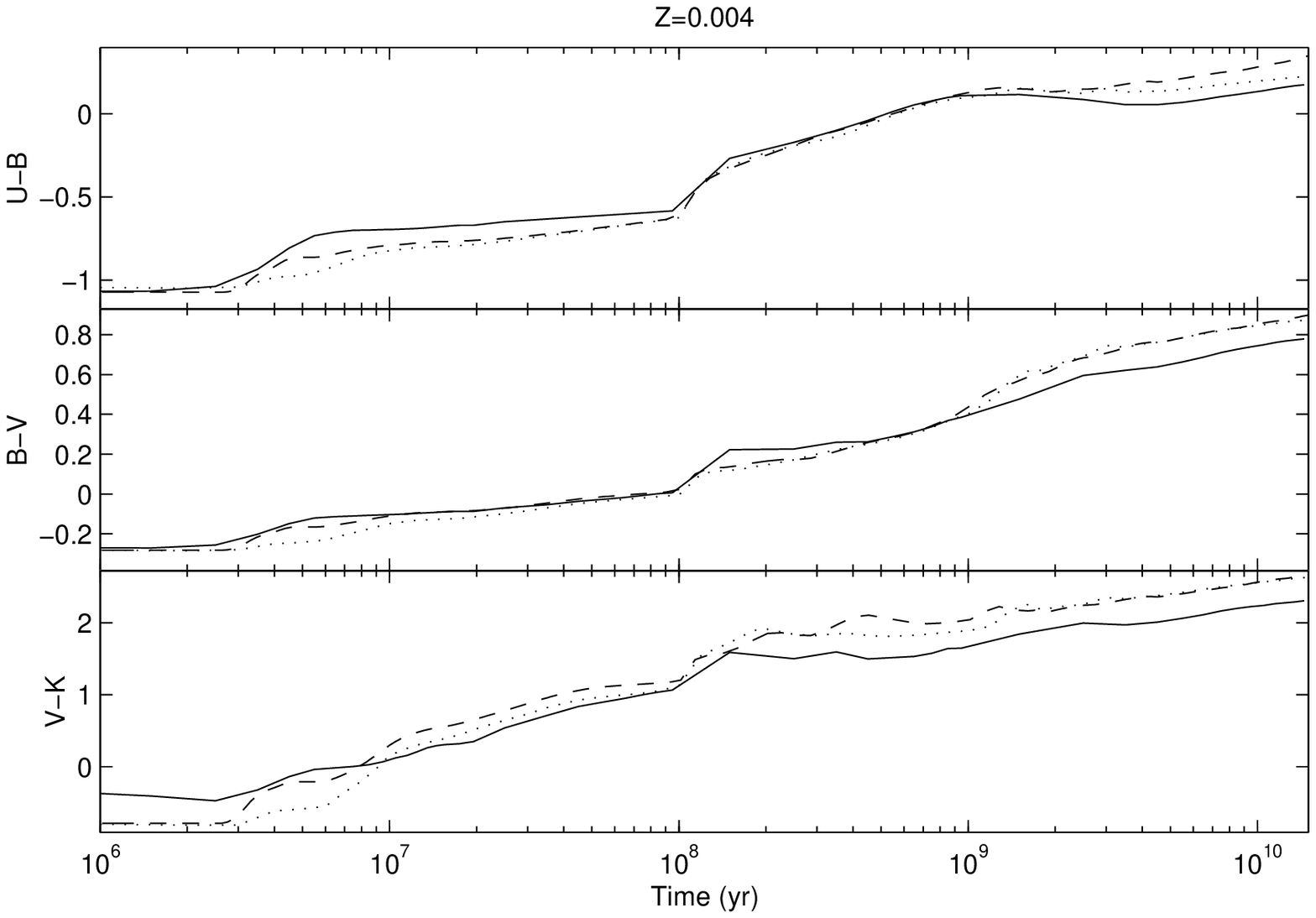}}
\resizebox{\hsize}{!}{\includegraphics{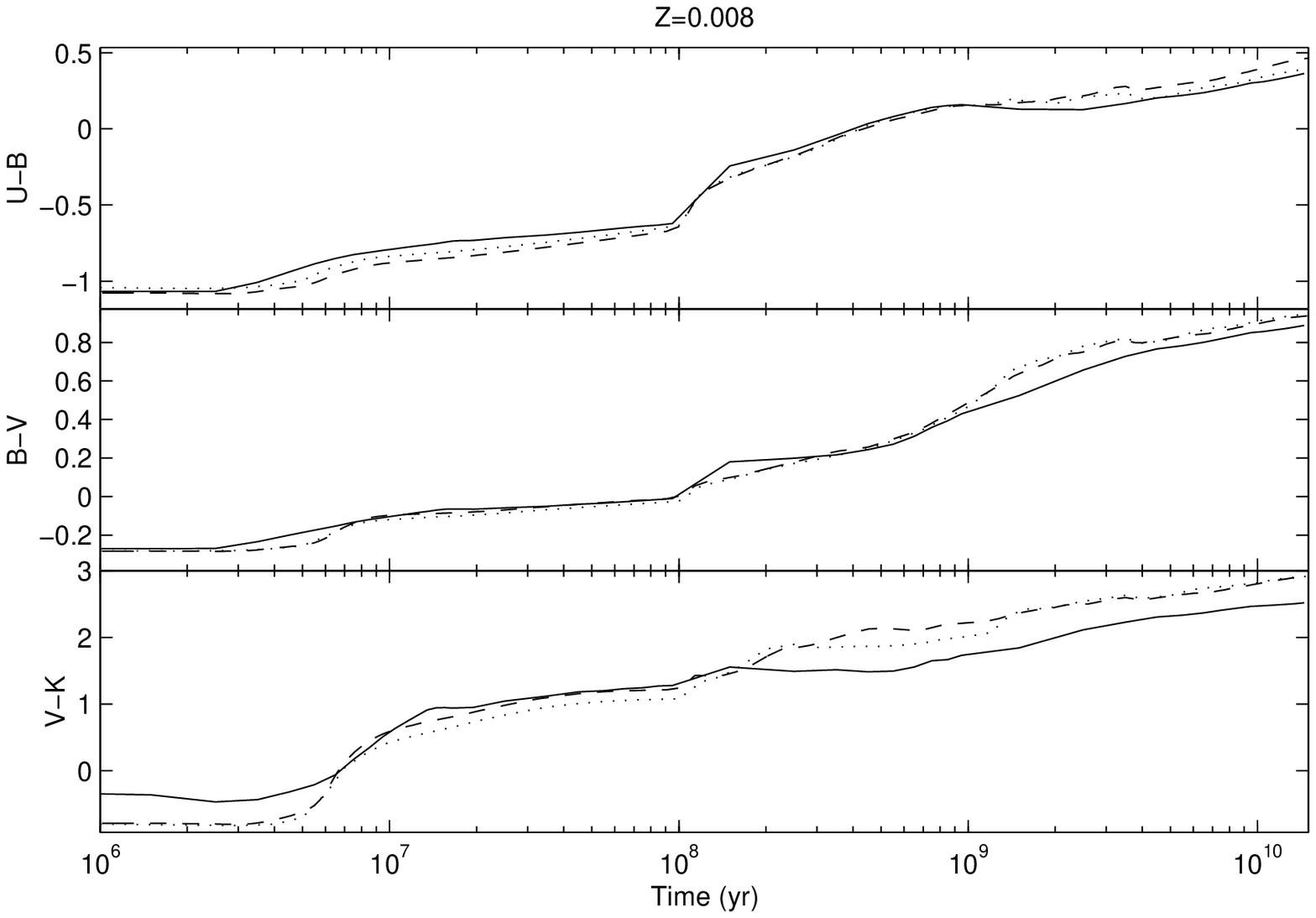}}
\resizebox{\hsize}{!}{\includegraphics{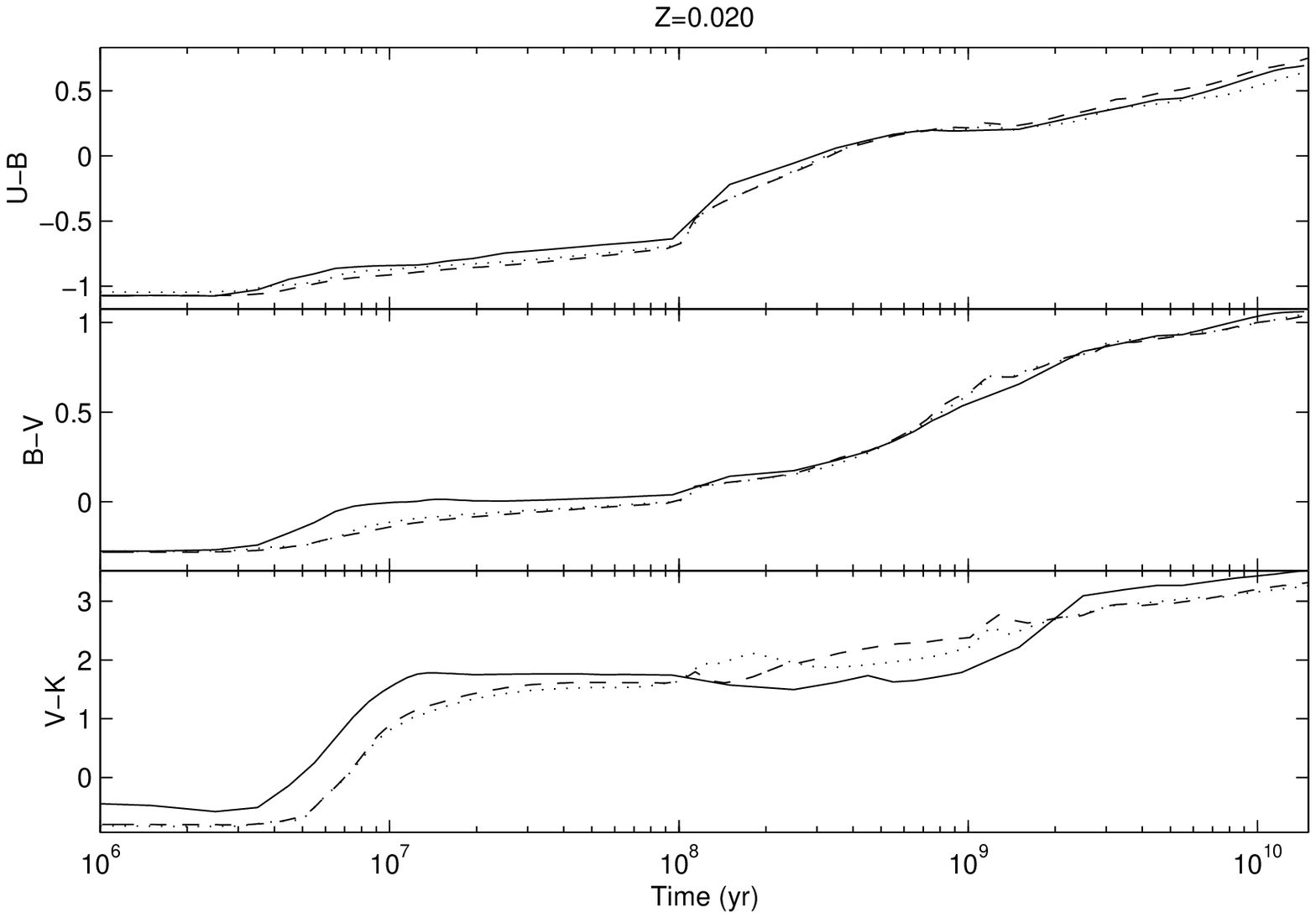}}
\caption[]{The evolution of the $U-B$, $B-V$ and $V-K$ colours predicted by our code (solid), BC96 (dashed) and PEGASE2 (dotted) at $Z$=0.004, 0.008 and 0.020. All models have been computed assuming a Salpeter IMF (0.08-120 $M_\odot$ for our code, 0.1-120 $M_\odot$ for PEGASE2, 0.1-125 $M_\odot$ for BC96), a short burst of constant star formation ($\tau$=100 Myr) and no nebular contribution.}
\label{galpegbc_s}
\end{figure}

\begin{figure}[h]
\resizebox{\hsize}{!}{\includegraphics{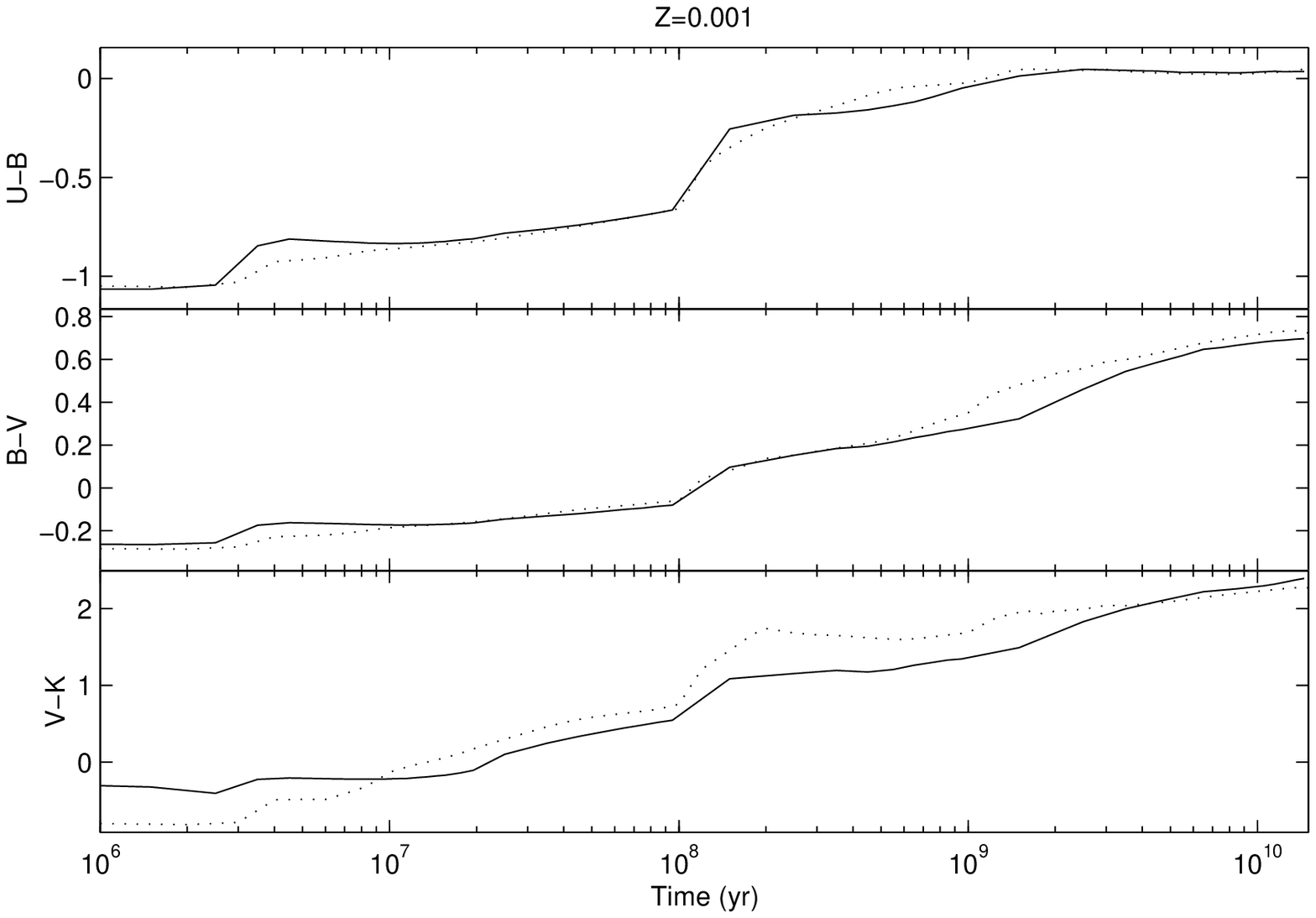}}
\resizebox{\hsize}{!}{\includegraphics{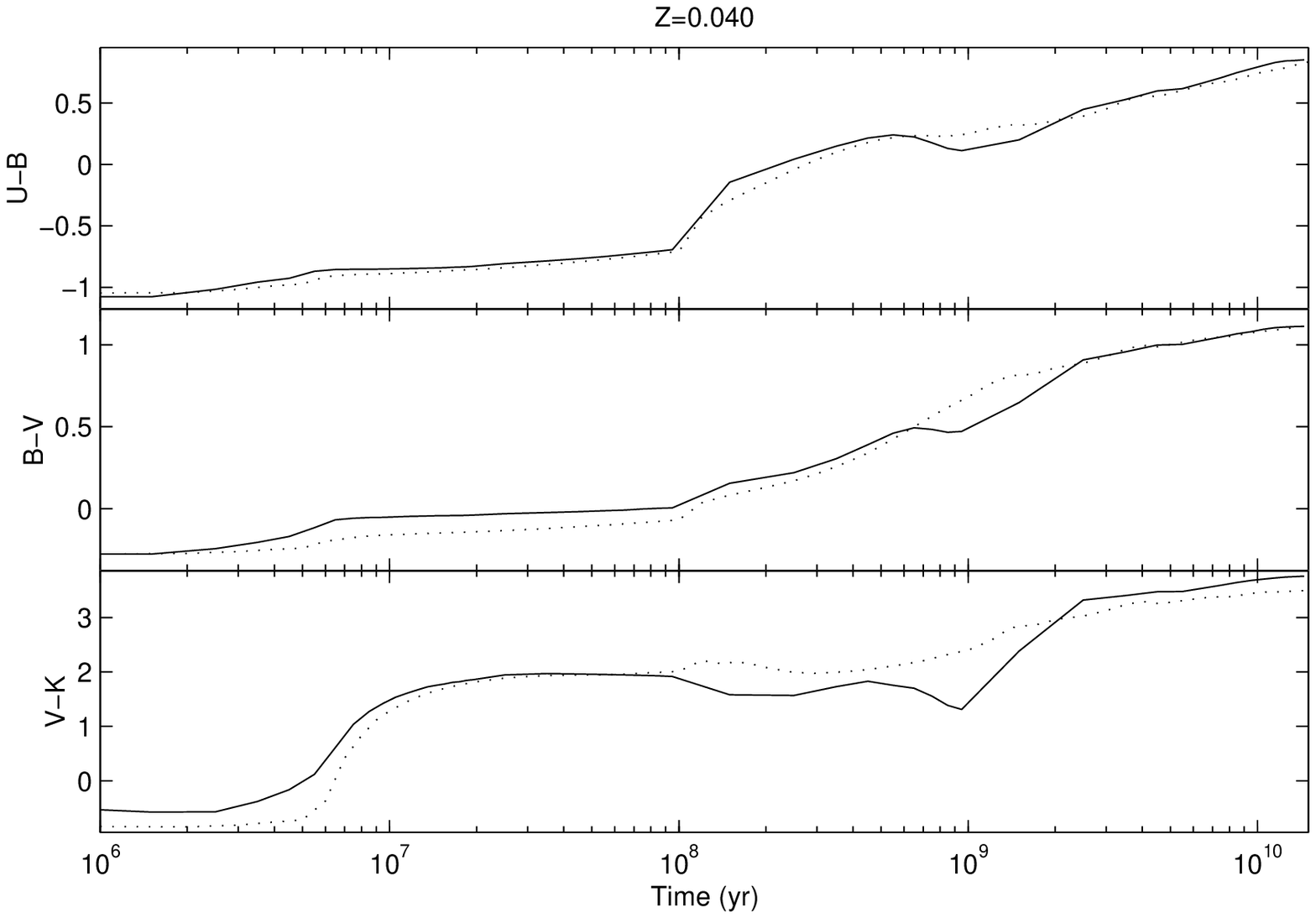}}
\caption[]{The evolution of the $U-B$, $B-V$ and $V-K$ colours predicted by our code (solid) and PEGASE2 (dotted) at $Z$=0.001 and 0.040. Otherwise same as  Fig.~\ref{galpegbc_s}.}
\label{galpeg_s}
\end{figure}

In Fig.~\ref{galpegbc_s} the photometric evolution predicted at $Z$=0.004, 0.008 and 0.008 for a pure stellar population undergoing a short burst of star formation is plotted for all three codes. This scenario is more suited for a quantitative comparison than any of the more continuous star formation histories, since a larger range of masses will be able to contribute to the integrated colours. Fig.~\ref{galpeg_s} contains the same comparison at $Z$=0.001 and $Z$=0.040 for our code and PEGASE2. 

Our code, PEGASE and BC96 all seem to agree fairly well in $U-B$ and $B-V$ colours, with differences peaking at 0.2 magnitudes. However, in $V-K$ the differences may reach as high as 0.7 magnitudes. Table 4 lists the time-averaged discrepancies between the three models along the colour sequences of Fig.~\ref{galpegbc_s}. BC96 and PEGASE2 generally seem to be in better agreement with each other than with our model. This is hardly surprising, since our model is mainly based on tracks from the Geneva set, whereas BC96 and PEGASE2 both employ stellar evolutionary tracks from the Padova set. Only in $U-B$ does the mean difference between our model and PEGASE2 become smaller than that of BC96 and PEGASE2. Apart from the Geneva-Padova discrepancy, the pronounced differences in $V-K$ between our model and the other two are enhanced by our use of evolutionary stages and strategies not adopted in these codes. The inclusion of PMS makes $V-K$ redder during the first few Myr of evolution, whereas the bluer $V-K$ seen in the age interval 0.1-1 Gyr is likely due to the different prescriptions used for the TP-AGB (Fioc \& Rocca-Volmerange \cite{Fioc & Rocca-Volmerange}). Finally, the effects of our extended HB morphologies at the three lowest metallicities become important after about 1 Gyr, producing significantly bluer $V-K$ at $Z$=0.004 and $Z$=0.008. The slightly redder $V-K$ predicted at $Z$=0.001 indicates a decreasing sensitivity to the HB morphology at this metallicity and the dominance of other evolutionary stages. Unfortunately, the discrepancies stemming from the different approaches to the TP-AGB and second parameter problem can not easily be disentangled from the different stellar tracks, atmospheres and computational methods used. To properly resolve this issue would require a much more careful analysis that is not within the scope of this paper.


\begin{figure}[h]
\resizebox{\hsize}{!}{\includegraphics{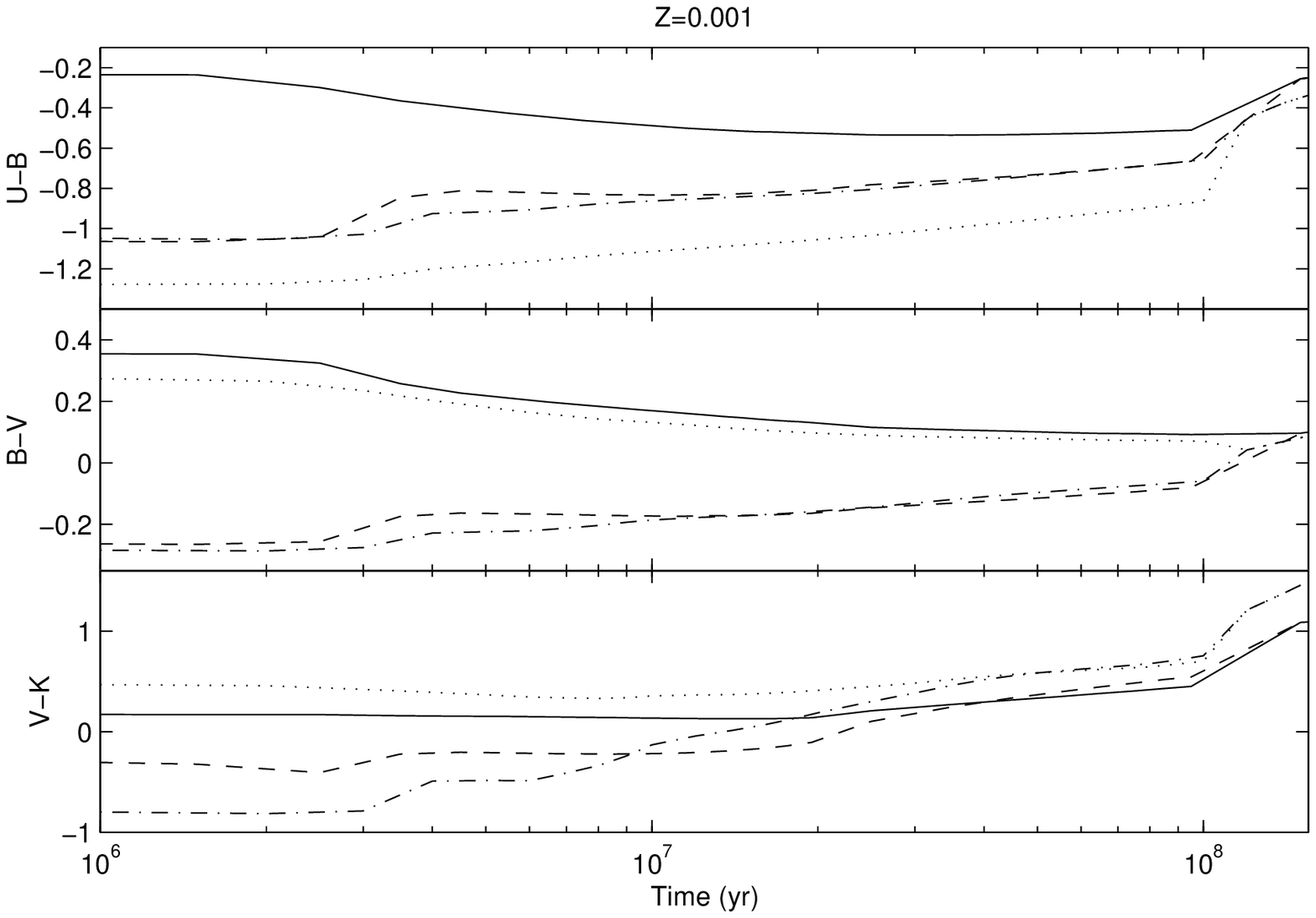}}
\resizebox{\hsize}{!}{\includegraphics{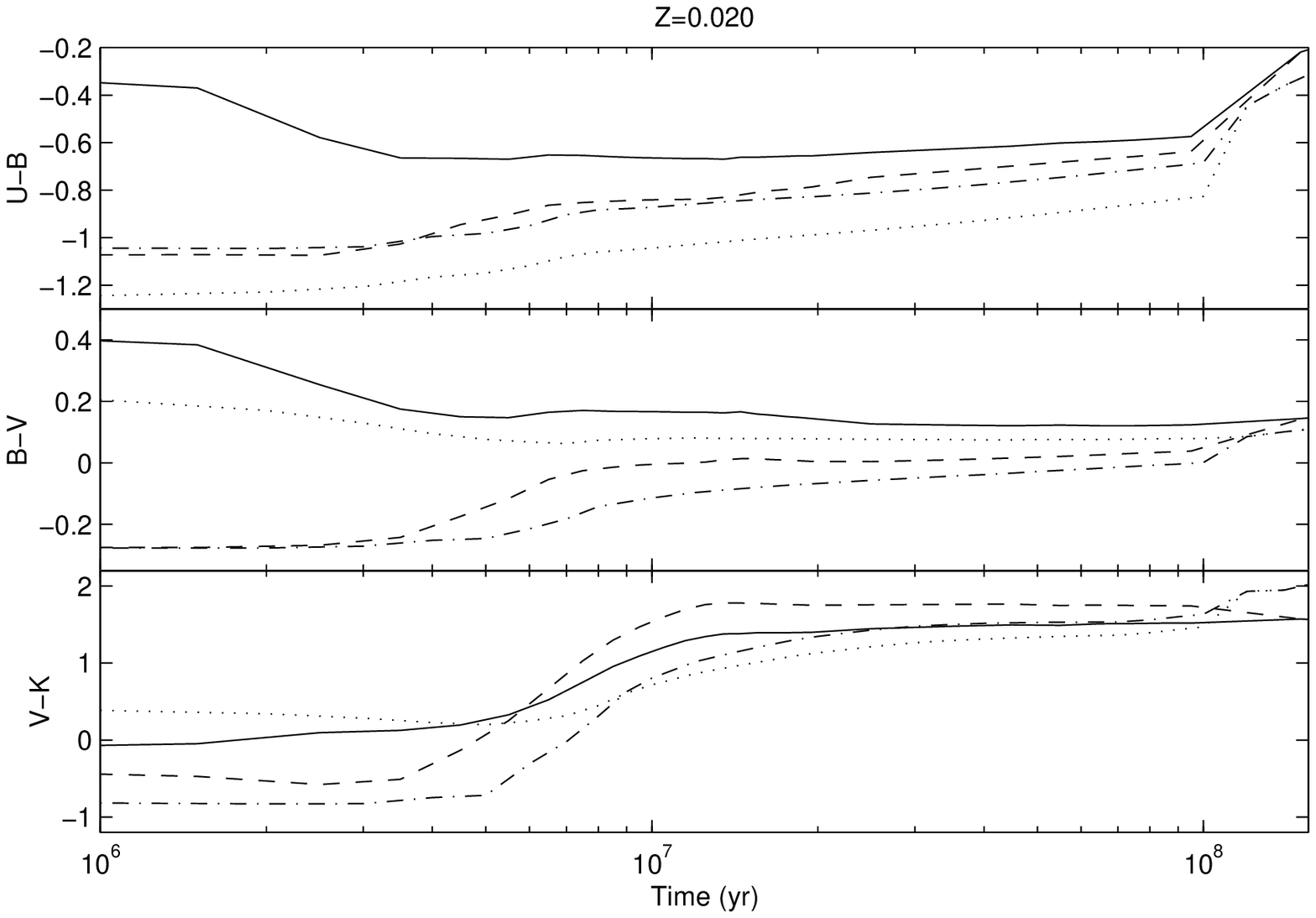}}
\caption[]{The evolution of the $U-B$, $B-V$, $V-K$ colours predicted with/without the inclusion of the nebular component by our code (solid/dashed) and PEGASE2 (dotted/dash-dotted) at $Z$=0.001 and $Z$=0.020, assuming a short burst of constant SFR ($\tau$=100 Myr), a Salpeter IMF (0.08-120 $M_\odot$ for our code, 0.1-120 $M_\odot$ for PEGASE).}
\label{galpeg_s_and_n}
\end{figure}

In Fig.~\ref{galpeg_s_and_n} we compare the photometric evolution predicted for stellar populations experiencing a short, constant burst of star formation at $Z$=0.001 and $Z$=0.020 using our code and PEGASE2, with and without the inclusion of the nebular component. Since the input parameters used by our photoionization model cannot easily be translated into the language of PEGASE2, nebular contribution in PEGASE2 was computed using default settings, while we preferred the use of $n$(H)=100 $\mathrm{cm^{-3}}$, a filling factor of 1.0, a covering factor of unity and $10^{10} \ M_\odot$ available for star formation. 
Even though both models start out with identical metallicities ($Z\mathrm{_{stars}}$=$Z\mathrm{_{gas}}$), evolution in the gaseous metallicity of PEGASE2 cannot not be prevented, making the comparison less adequate at high ages. In the evolutionary sequences used here, PEGASE2 will reach $Z\mathrm{_{gas}}$=0.004 at $t\approx 95$ Myr in the $Z\mathrm{_{stars}}$=0.001 scenario and $Z\mathrm{_{gas}}$=0.040 at $t\approx 75$ Myr when $Z\mathrm{_{stars}}$=0.020.
  
As seen in Fig.~\ref{galpeg_s_and_n}, properly taking the effects of nebular emission into account becomes crucial during the star-forming phase. The differences in colour between scenarios with and without nebular emission peak at 0.8 (0.2) in $U-B$ for our code (PEGASE2), 0.7 (0.6) in $B-V$ and 0.7 (1.3) in $V-K$. Due to the different input parameters, stellar SEDs and photoionization models used the final colours are however very different, especially in $U-B$, where the sequences actually shift in different directions when nebular emission is added. The significantly bluer $U-B$ produced by PEGASE2 appears to stem from a much higher luminosity predicted for the [OII]$\lambda$3727 emission line throughout the entire star formation episode ([OII]$\lambda$3727/H$\beta \approx 3$ already at $t$=1 Myr). Even though a direct comparison of these colours to those seen in starburst galaxies may be complicated by the possible presence of dust, it should be noted that colours as blue as the ones predicted by PEGASE2 are not typically observed. Out of the $\sim 30$ blue compact galaxies (assumed to have $U-B < -0.1$) in the sample of Bergvall \& Olofsson (\cite{Bergvall & Olofsson}) and 17 blue compact dwarf galaxies (for which this colour was available) in Thuan (\cite{Thuan}), none have $U-B < -1$.

Since reliable tests of SEMs are yet to be developed, there is no simple way around these model discrepancies. Our recommendation would therefore be to conduct multi-model consistency checks of all important conclusions whenever comparable codes are available.

\section{Comparison to observations}
To check the performance of our model in confrontation with reality, we have made a comparison of predicted broadband colours to those of observed globular clusters (GCs) and low-surface brightness galaxies (LSBG). Predicted line ratios and equivalent widths have been compared to observed HII galaxies. Unless otherwise stated, all models assume $Z_{\mathrm{gas}}$=$Z_{\mathrm{stars}}$, covering factor 1.0, filling factor 1.0, $n(\mathrm{H})$=100 $\mathrm{cm^{-3}}$ and a Salpeter IMF (0.08-120 $M_\odot$).

\subsection{Broadband colours}
\begin{figure}[t]
\resizebox{\hsize}{!}{\includegraphics{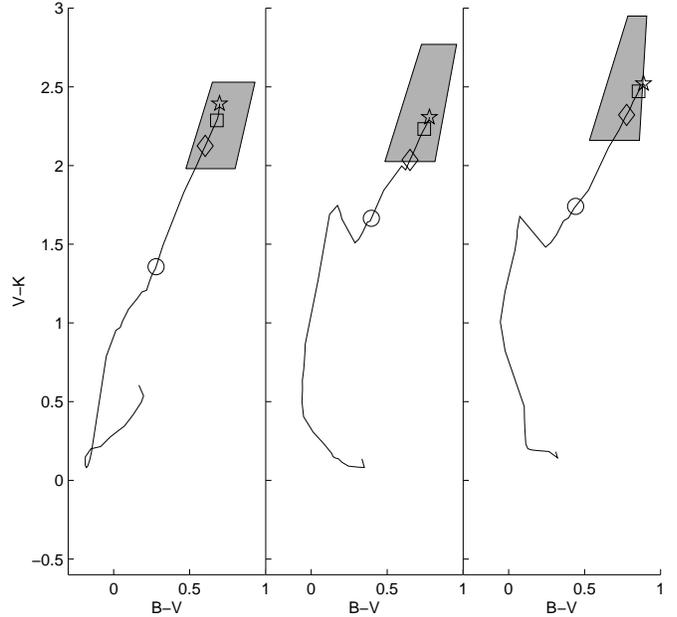}}
\caption[]{Regions in $B-V$ vs $V-K$ space inhabited by confirmed GCs in M31, subject to metallicity binning in $V-K$. Left: Evolutionary sequence (solid) predicted at $Z$=0.001 and the region inhabited by objects in the metallicity range $Z$=0.0005-0.002 (shaded area). Middle: Evolutionary sequence at $Z$=0.004 and objects in the range $Z$=0.002-0.006. Right: Evolutionary sequence at $Z$=0.008 and objects in the range $Z$=0.006-0.014. All models assume a Salpeter IMF (0.08-120 $M_\odot$, a short burst of constant star formation ($\tau$=10 Myr) and $10^{5} \ M_\odot$ available for star formation. For each model sequence, circles mark the predicted colours at 1 Gyr, diamonds at 5 Gyr, squares at 10 Gyr and pentagrams at 15 Gyr.}
\label{GCs_gal_multiz}
\end{figure}

Even though our model, due to the statistical treatment of HB morphologies, cannot be expected to reproduce the average colours of low-metallicity globular clusters, as discussed in section 2.3, we would expect the predicted colours to fall somewhere within the range observed. In order to test this property, we have compared our evolutionary sequences to the large set of GCs in M31, as compiled by Barmby et al. (\cite{Barmby et al.}). Since the metallicities and dereddened colours derived for theses objects are only available in figures, the comparison is carried out with respect to the approximate outer boundary of the region inhabited by the GCs the $B-V$, $V-K$ two-colour diagram, especially suited to test the treatment of late stages of stellar evolution. The boundaries have been subjected to metallicity binning in $V-K$ but not in $B-V$. Since our main interest is to make sure that the extended HB morphologies produced by our approach to the second parameter problem do not produce sequences that are too blue in $V-K$ at high ages, this is of no major concern. As seen in Fig.~\ref{GCs_gal_multiz}, the evolutionary sequences at $Z$=0.001, 0.004 and 0.008 all fall well within the range of colours seen in this sample for ages higher than about 5 Gyr.
\begin{figure}[h]
\resizebox{\hsize}{!}{\includegraphics{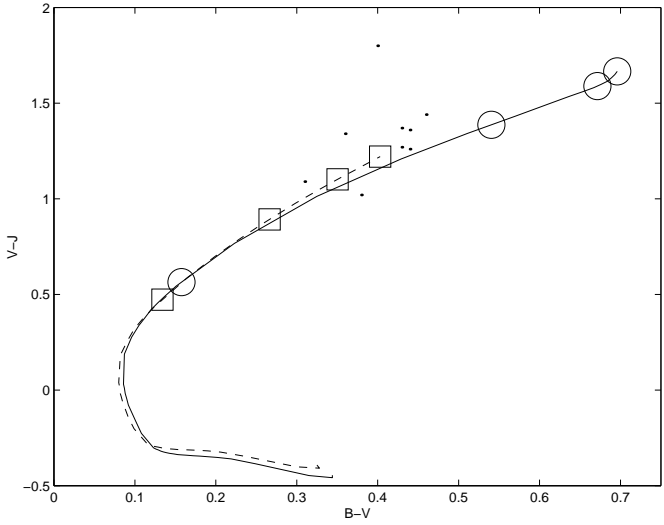}}
\caption[]{$B-V$ vs $V-J$ for 9 blue LSBGs (dots). Lines are evolutionary sequences predicted at $Z$=0.001 for exponentially declining SFRs with $\tau$=1 Gyr (solid) and $\tau$=14 Gyr (dashed), assuming $10^{10} \ M_\odot$ available for star formation. Along each sequence, ages of 1, 5, 10 and 15 Gyr have been illustrated with markers (circles for solid, squares for dashed). }
\label{LSBGs_z001}
\end{figure}

We now turn our attention to a set of blue LSBGs (Bergvall et al. \cite {Bergvall et al. II}; R\"onnback \& Bergvall \cite{Rönnback & Bergvall}) which despite being actively star-forming and metal-poor are believed to have ages higher than 2 Gyr (Bergvall \& R\"onnback \cite{Bergvall & Rönnback}). Standard models of chemical evolution cannot easily account for this phenomenon without fine-tuning the unknown rates of infall to and mass-loss from these galaxies. A constant metallicity model may therefore be just as appropriate for describing their evolution. A detailed analysis of the star formation history of these objects is however not within the scope of this paper. Here, we are simply want to  establish that the colours observed can be reasonably well reproduced by our model. The low level of dust extinction observed in these objects allows the predicted colours to be directly compared to the observations without any correction for reddening. In Fig.~\ref{LSBGs_z001} we compare the location of the observed LSBGs in the $B-V$, $V-J$ two-colour diagram to the evolutionary sequences predicted for scenarios believed to span the likely range of present star formation in this sample. With the exception of one object, separated from the others in $V-J$, the model appears quite compatible with the colours observed. Even though the scenario with shorter time-scale of star formation probably underestimates the current SFR in these objects, it places firm constraints on the lowest ages allowed, indicating - in agreement with previous findings - that these galaxies cannot be young.     
 
\subsection{Line ratios and equivalent widths} 
For the spectroscopic data, we have been using the compilation of HII galaxies by Masegosa et al. (\cite{Masegosa et al.}), consisting of 121 objects, all corrected for intrinsic reddening. This data-set was also analyzed by Stasi\'nska \& Leitherer (\cite{Stasinska & Leitherer}), which allows the performance of their model to be directly compared to ours. 

Since the starburst regions sampled in this data-set are of varying spatial extent, but typically smaller than the host HII galaxy, the total mass available for star formation has been considered a free parameter. We have furthermore assumed that the contribution to the SED from the older, underlying stellar component is insignificant.  

\begin{figure}[t]
\resizebox{\hsize}{!}{\includegraphics{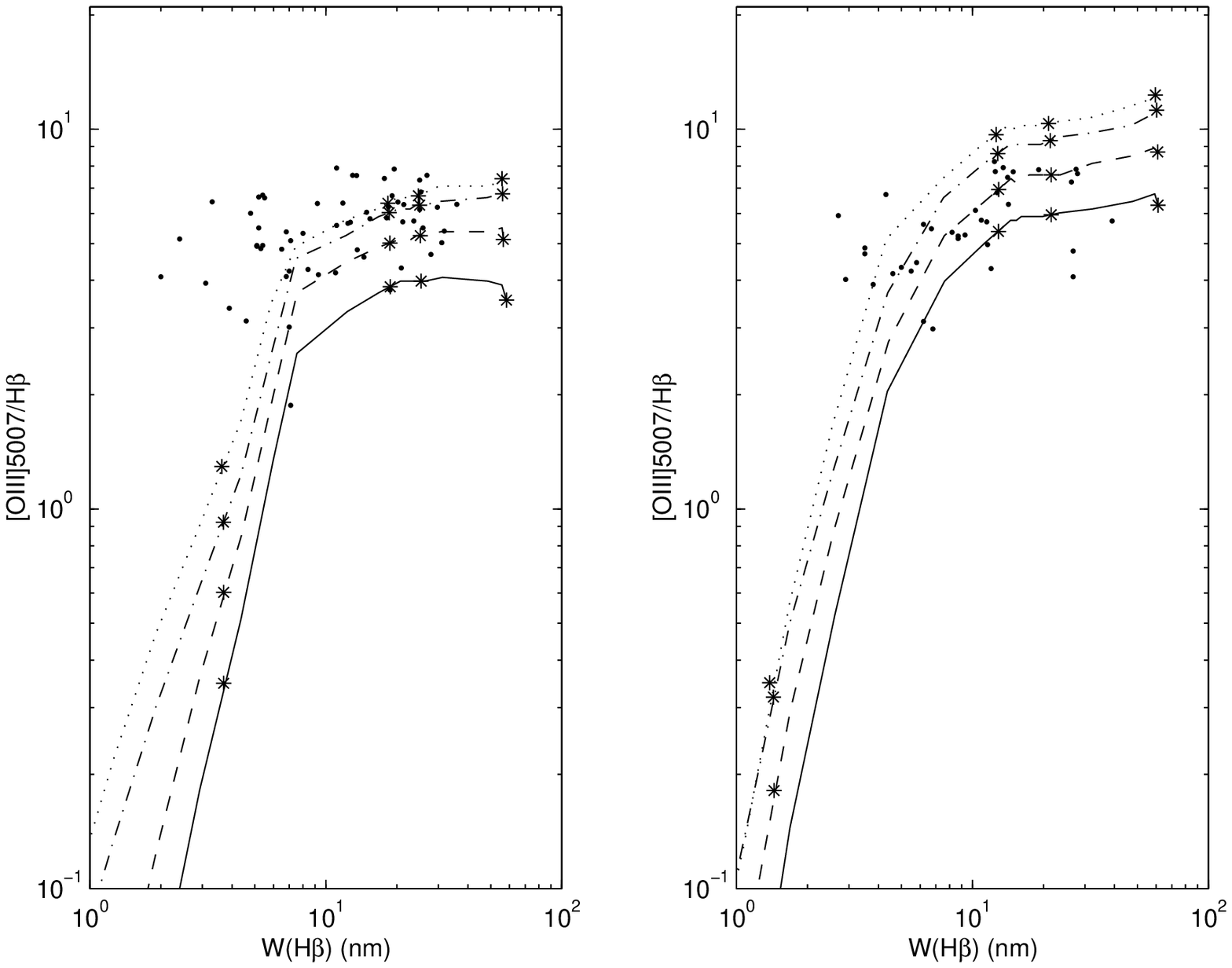}}
\resizebox{\hsize}{!}{\includegraphics{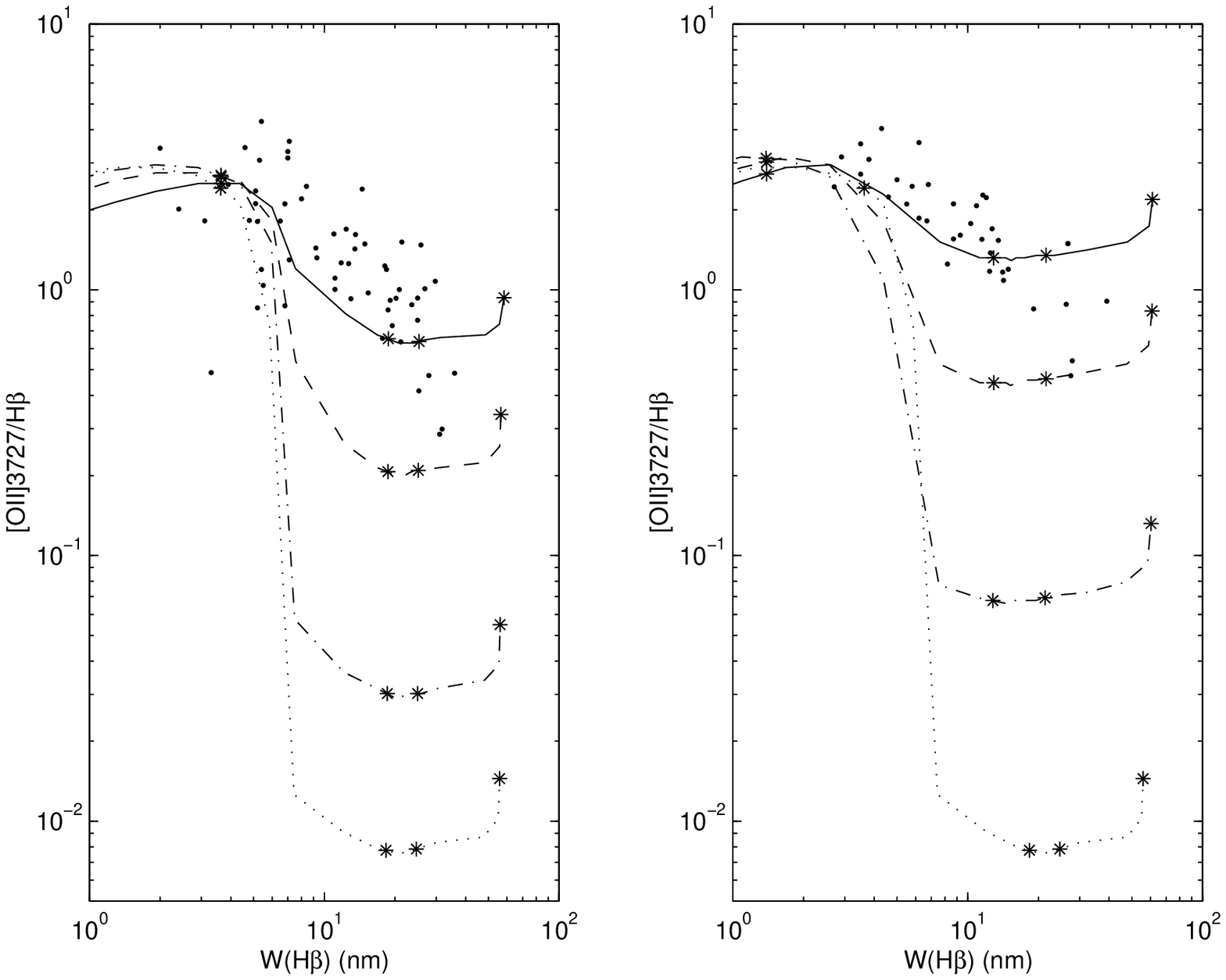}}
\caption[]{Evolutionary sequences of [OIII]$\lambda$5007/H$\beta$ vs $W$(H$\beta$) and [OII]$\lambda$3727/H$\beta$ vs $W$(H$\beta$) assuming $10^3 \ M_\odot$ (solid), $10^5 \ M_\odot$ (dashed), $10^{8} \ M_\odot$ (dash-dotted) and $10^{10} \ M_\odot$ (dotted) available for star formation. All models use a Salpeter IMF (0.08-120 $M_\odot$) and a short burst of constant SFR ($\tau$=10 Myr). The observed sample of HII galaxies is represented by  dots. Asterisks mark ages of 0.5, 5, 10 and 15 Myr. Left: Evolutionary sequences assuming $Z=0.001$ and observed metallicities in the range $Z$=0.0005-0.002. Right: Evolutionary sequences assuming $Z$=0.004 and observed metallicities in the range $Z$=0.002-0.006.}
\label{lines_obscomp}
\end{figure}

Fig. \ref{lines_obscomp} presents observations and evolutionary sequences in two diagnostic diagrams, [OIII]$\lambda$5007/H$\beta$ vs $W$(H$\beta$) and [OII]$\lambda$3727/H$\beta$ vs $W$(H$\beta$). The evolutionary sequences of [OIII]$\lambda$5007/H$\beta$ vs $W$(H$\beta$) predicted for starburst masses in the range $10^5-10^{10} \ M_\odot$ all seem to agree reasonably well with the observations. The value of $W$(H$\beta$) at which the predicted [OIII]$\lambda$5007/H$\beta$ starts to drop is sensitive to the duration of the starburst, and the value of [OIII]$\lambda$5007/H$\beta$ for objects located at $W$(H$\beta$) lower than 5 nm may indicate more prolonged star formation scenarios than the one assumed here.

\begin{figure}[t]
\resizebox{\hsize}{!}{\includegraphics{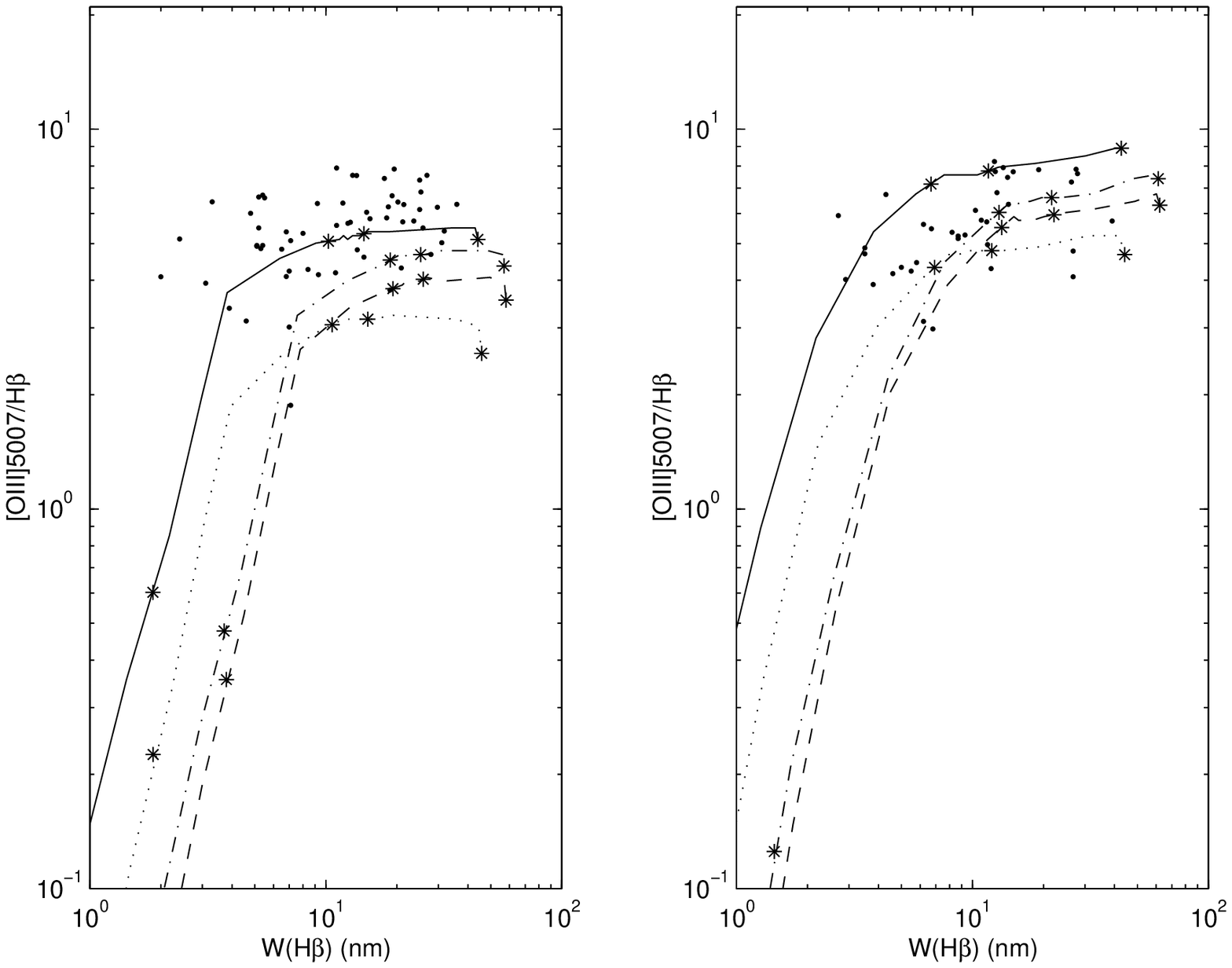}}
\resizebox{\hsize}{!}{\includegraphics{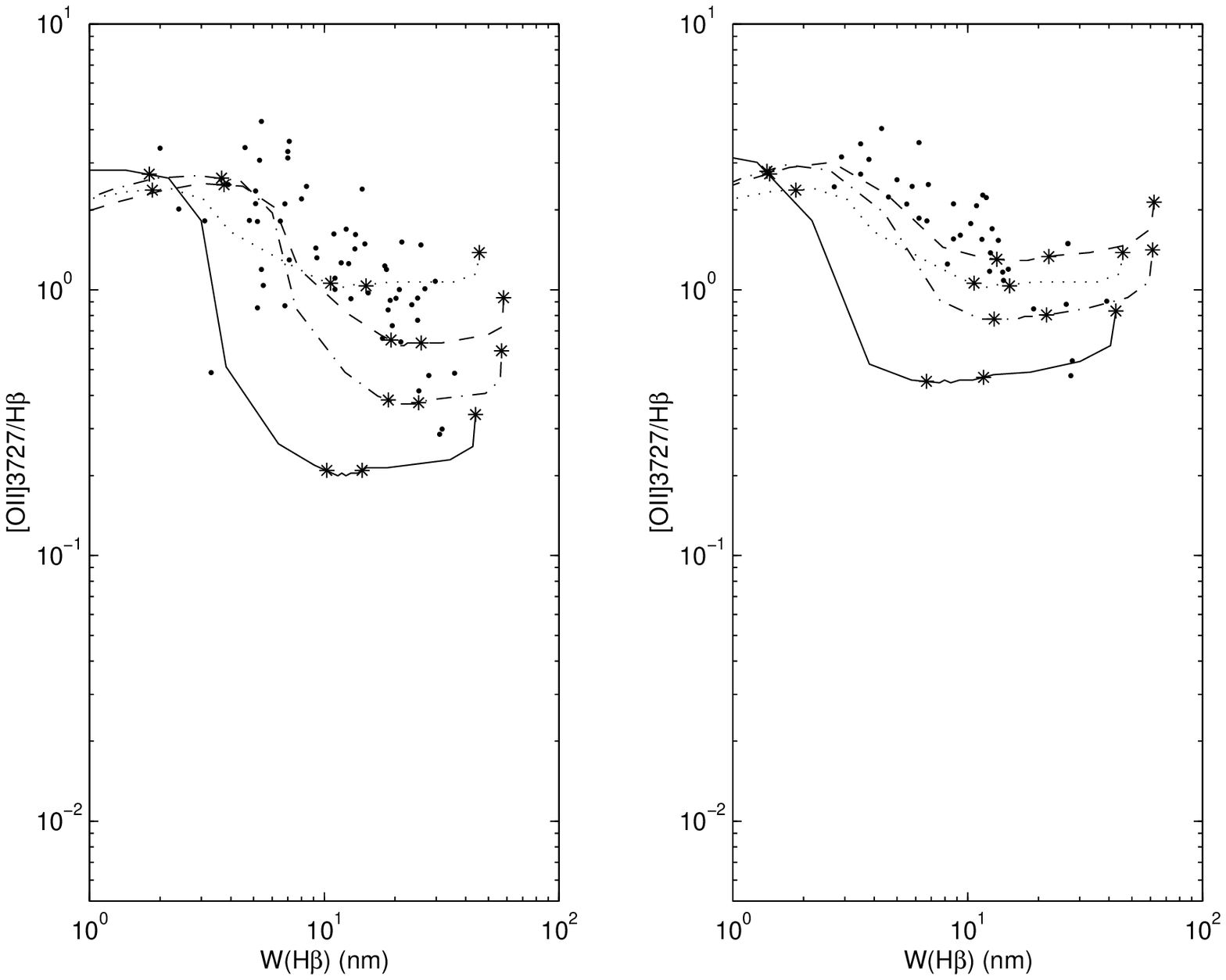}}
\caption[]{The gas parameter dependence of the $10^5 \ M_\odot$ model from Fig. \ref{lines_obscomp}. The covering factor, filling factor and hydrogen number density ($\mathrm{cm^{-3}}$) were allowed to take on the following values: 0.5, 1.0, 100 (solid); 1.0, 0.1, 100 (dashed); 1.0, 1.0, 10 (dash-dotted); 0.5, 0.1, 10 (dotted).}
\label{lines_obscomp_paramtest}
\end{figure}

The [OII]$\lambda$3727/H$\beta$ vs $W$(H$\beta$) diagram presents a much more troublesome picture. Neither the slope nor the absolute values of the observed [OII]$\lambda$3727/H$\beta$-sequence can easily be reproduced by our model. The line ratio is significantly underpredicted during the entire duration of the starburst for all but the lowest mass model, which performs poorly in [OIII]$\lambda$5007/H$\beta$ vs $W$(H$\beta$), especially in the lower metallicity bin. A similar trend is seen in Stasi\'nska \& Leitherer, where only the $10^3 \ M_\odot$ model - also rating poorly in [OIII]$\lambda$5007/H$\beta$ vs $W$(H$\beta$) - was able to reproduce the high values of [OII]$\lambda$3727/H$\beta$. Assuming masses and efficiencies typical of star-forming regions, these low-mass models appear quite unlikely. It seems reasonable to believe that the stellar mass in this sample should be at least $10^5$-$10^6$ $M_\odot$. In Fig.~\ref{lines_obscomp_paramtest}, we therefore investigate the dependence of the $10^5 \ M_\odot$ model on different gas parameters, and show that the discrepancies between the performance in the two diagnostic diagrams cannot easily be resolved this way. Since this curiosity has now been confirmed by two independent codes, using different stellar evolutionary tracks, stellar atmospheres and photoionization models, this may indicate something missing from our understanding of the physics of these objects. Other observational studies of emission line galaxies (e.g. Kniazev et al. \cite{Kniazev et al.}) show similar trends in these diagnostic diagrams, indicating that severe problems with the actual data-set can probably be ruled out. A plausible explanation for the discrepancy could however be the existence of spatial variations in the physical conditions (e.g. density and filling factor) between the [OIII] and [OII] regions of the nebula. There could also be a mixture of several distinct HII regions with different characteristics sampled by the slit of the spectrograph, thereby making the model assumption of a single, homogeneous HII region inappropriate. Part of the nebular emission may furthermore originate from regions ionized by shocks or non-thermal sources. As a test of this last possibility, we have compared the observed emission line ratios to those predicted by a simple model using a mixture of photoionized, shocked and power-law ionized regions. Three regions were mixed, either 3 HII, 2 HII + 1 shock, 2 HII+1 power-law or
1 HII + 1 shock + 1 power-law region. Details about the procedure may be found in Bergvall et al. (\cite{Bergvall et al. I}). In $\approx$ 20\% of objects, combinations including power-laws or shocks gave significantly better fits than did pure mixtures of HII regions. This indicates that shocks and non-thermal sources could, to some degree, influence line ratios in this sample and explain part of the consistency problems between models and observations.

\section{Summary and conclusions}
We have presented the first version of a new spectral evolutionary code for galaxies in the age range 0-15 Gyr. The model represents an improvement with regard to existing codes in the inclusion of PMS evolution and a sophisticated nebular component.  

The time durations of significant nebular emission in different photometric filters are estimated for several star formation scenarios and metallicities. These estimates may serve as a guide to the age at which the common assumption of negligible nebular emission in early-type galaxies breaks down. The sensitivity of the colours predicted to the physical properties of gas (i.e hydrogen density, metallicity, covering and filling factors) is also tested, revealing a high sensitivity to the metallicity and covering factor, and a lower, yet observationally significant, sensitivity to the density and filling factor assumed. This has implications for the modelling and analysis of young galaxies, where normally gas parameter values only typical of star-forming regions are applied, and stresses the need for a more careful analysis of the possible parameter space.

The predicted evolution of the H$\alpha$, H$\beta$ and [OIII]$\lambda$5007 line equivalent widths is presented for different metallicities and values of gas parameters. The H$\alpha$, H$\beta$ equivalent widths are found only to be sensitive to the covering factor whereas [OIII]$\lambda$5007 appears sensitive to the choice of other parameters as well. It is noted that the oscillating and increasing equivalent width of [OIII]$\lambda$5007 predicted by Stasi\'nska \& Leitherer (\cite{Stasinska & Leitherer}) at solar metallicity, thereby hampering the use of this line as an age indicator, is not reproduced by our model. 
 
The contribution from the PMS phase to the integrated properties of a young stellar populations is evaluated and found to be substantial in the near-IR during the first few million years of a standard IMF starburst lasting 100 Myr, e.g. increasing the luminosity in the K-band by as much as 0.34 magnitudes at $Z$=0.001 and 0.16 at $Z$=0.020 when compared to the case where all stars are assumed to simultaneously start on the ZAMS at time equal to zero. If more extreme IMFs are to be modelled, the effects become significant in all filters and for much more prolonged periods of time. 

A comparison to the colour predictions of the stellar components from two similar models, BC96 and PEGASE2, reveal a number of discrepancies between the codes, especially in the near-IR, where predictions are highly affected by uncertainties in the treatment of late stages of stellar evolution. Since reliable tests of SEMs are yet to be developed, this result emphasizes the need for multi-model consistency checks whenever comparable codes are available.

Finally, the colours predicted by our model are tested against observed samples of GCs and LSBGs, showing an acceptable agreement. Line ratios and equivalent widths are then compared to a set of HII galaxies. Even though the model performs well in the [OIII]$\lambda$5007/H$\beta$ vs $W$(H$\beta$) diagnostic diagram, the difficulty in simultaneously fitting [OIII]$\lambda$5007/H$\beta$ vs $W$(H$\beta$) and [OII]$\lambda$3727/H$\beta$ vs $W$(H$\beta$) noted in Stasi\'nska \& Leitherer (\cite{Stasinska & Leitherer}) does however persist in our study as well. This may indicate something missing from our understanding of the physics of these objects. 

We conclude that our model is appropriate for the modelling of young as well as old galaxies in the low- to intermediate mass range. Due to the large uncertainties still prevailing the field of spectral evolutionary synthesis (e.g. the near-IR evolution of stellar populations, the colours produced by inclusion of the nebular component), we do not recommend the use of any model by itself in the analysis of galaxies, and stress that all important conclusions should be cross-checked with other codes as well.

\begin{acknowledgements}
We wish to thank Ken Mattsson for substantial contributions to the coding of the model and Arnaud Pharasyn for developing a powerful graphical interface for visualization of the SEDs. This project was partly supported by the Swedish Space Board.  
\end{acknowledgements}

\end{document}